\documentclass[12pt]{spieman}  % 12pt font required by SPIE;
\usepackage{amsmath,amsfonts,amssymb}
\usepackage{graphicx}
\usepackage{setspace}
\usepackage{tocloft}
\usepackage{lineno}
\usepackage{siunitx}%%
\usepackage{rotating} % for sideways figure
\usepackage{xr} % for external references
\title{
    AI-Augmented Photon-Trapping Spectrometer-on-a-Chip on Silicon Platform with Extended Near-Infrared Sensitivity
}

\author[a]{Ahasan Ahamed}
\author[a]{Htet Myat}
\author[a]{Amita Rawat}
\author[a]{Lisa N McPhillips}
\author[a*]{M Saif Islam}
\affil[]{University of California Davis, Electrical and Computer Engineering Department, One Shields Avenue, Davis, CA 95616, USA}

\renewcommand{\cftdotsep}{\cftnodots}
\cftpagenumbersoff{figure}
\cftpagenumbersoff{table} 
\begin{document} 
\maketitle

\begin{abstract}
    We present a compact, noise-resilient reconstructive spectrometer-on-a-chip that achieves high-resolution hyperspectral imaging across an extended near-infrared (NIR) range up to \SI{1100}{\nano\meter}. The device integrates monolithically fabricated silicon photodiodes enhanced with photon-trapping surface textures (PTST), enabling improved responsivity in the low-absorption NIR regime. Leveraging a fully connected neural network, we demonstrate accurate spectral reconstruction from only 16 uniquely engineered detectors, achieving $<$0.05 RMSE and $\sim$\SI{8}{\nano\meter} resolution over a wide spectral range of \SIrange{640}{1100}{\nano\meter}. Our system outperforms conventional spectrometers, maintaining signal-to-noise ratio above \SI{30}{\decibel} even with \SI{40}{\decibel} of added detector noise—extending functionality to longer wavelengths up to \SI{1100}{\nano\meter}, while the traditional spectrometers fail to perform beyond \SI{950}{\nano\meter} due to poor detector efficiency and noise performance. With a footprint of \SI{0.4}{\milli\meter\squared}, dynamic range of \SI{50}{\decibel}, ultrafast time response (\SI{57}{\pico\second}), and high photodiode gain ($>$7000), this AI-augmented silicon spectrometer is well-suited for portable, real-time, and low-light applications in biomedical imaging, environmental monitoring, and remote sensing. The results establish a pathway toward fully integrated, high-performance hyperspectral sensing in a CMOS-compatible platform.
\end{abstract}

% Include a list of up to six keywords after the abstract
\keywords{Hyperspectral imaging, Reconstructive Spectrometer, Near-infrared (NIR), Artificial Intelligence, On-chip spectrometer, Noise resilient}

% Include email contact information for corresponding author
{\noindent \footnotesize\textbf{*}M Saif Islam,  \linkable{sislam@ucdavis.edu} }

\begin{spacing}{2}   % use double spacing for rest of manuscript

\section{Introduction}
\label{Introduction}    % \label{} allows reference to this section
\begin{figure}[htbp]
\centering
\includegraphics[width=0.9\textwidth]{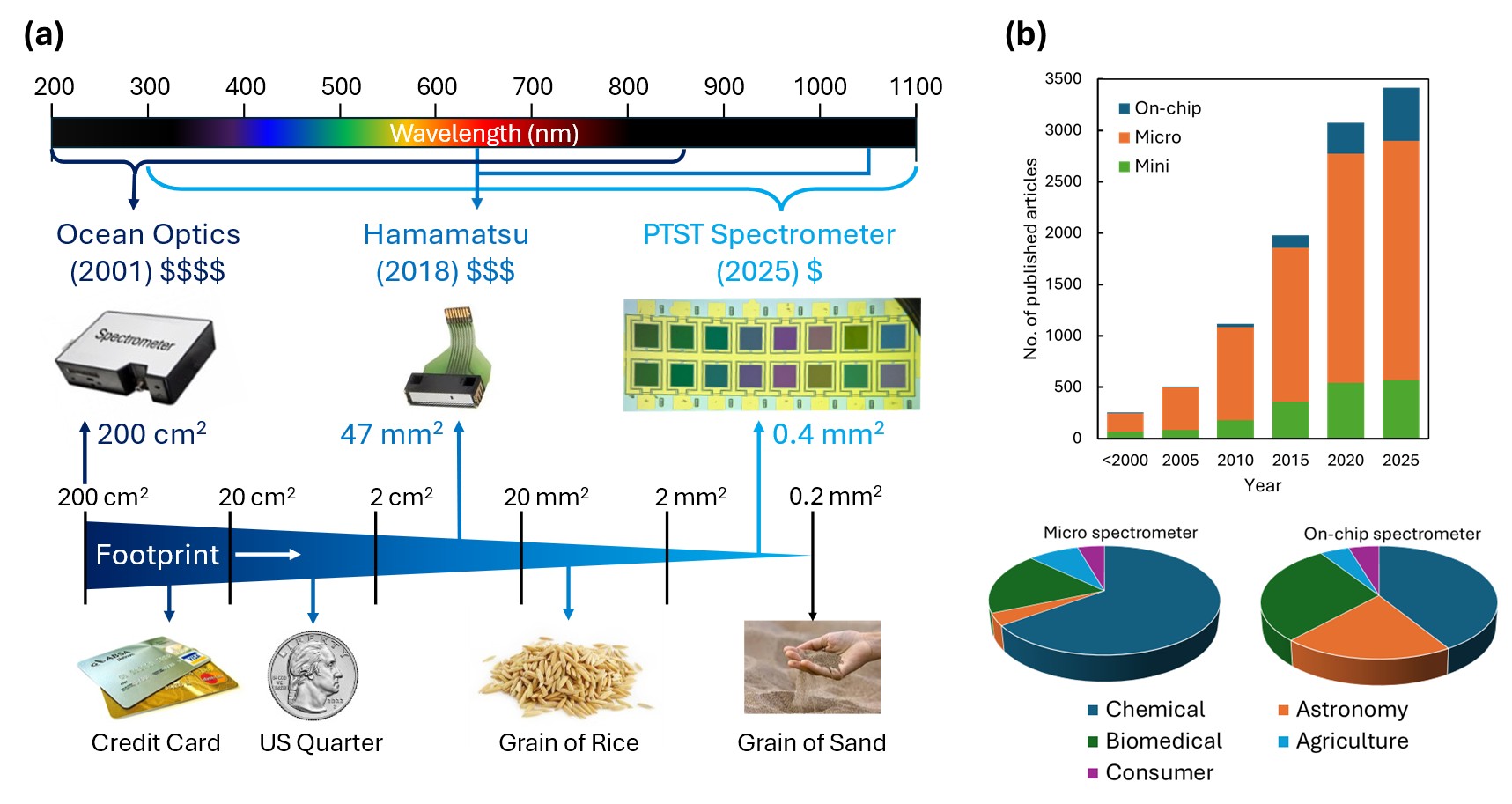}
\caption[Miniaturization trend of optical spectrometer]{Miniaturization trend of optical spectrometer: (a) From left to right, the figure illustrates the size reduction from bench-top to handheld to chip-scale spectrometers. (Left) A miniaturized commercial benchtop spectrometer with a footprint of \SI{200}{\centi\meter\squared} with spectral range from \SIrange{200}{850}{\nano\meter} \cite{oceanoptics2024USB4000}. (Center) A commercial handheld micro-spectrometer with a footprint of \SI{40}{\milli\meter\squared} with spectral range from \SIrange{640}{1050}{\nano\meter} \cite{hamamatsu2018microSpectrometer}. (Right) A chip-scale PTST spectrometer using an array of silicon photodiodes with a footprint of \SI{0.4}{\milli\meter\squared}, comparable to a grain of sand, sensitive over a broad spectral range of \SIrange{300}{1100}{\nano\meter} (this work). As the size decreases, the price of the spectrometers also drops significantly, much like the CMOS image sensors \cite{Fossum1995cmosImageSensors}.
(b) (Top) A chart showing the number of publications related to miniaturized spectrometers from the year 2000 to 2025 in 5-year intervals \cite{GoogleScholar}. A huge drive towards miniaturization of spectrometers has been observed in recent years. (Bottom) The distribution of potential applications for micro and on-chip spectrometers shows an increasing demand for on-chip spectrometers in biomedical imaging, astronomy, and consumer applications \cite{GoogleScholar}. 
}
\label{fig:miniaturization_trend}
\end{figure}

Hyperspectral imaging (HSI) is a ubiquitous technique that has emerged as a transformative tool for applications ranging from environmental monitoring and precision agriculture to biomedical diagnostics and industrial quality control \cite{Maldonado2018agriculturalHyperspectralImaging,Stuart2019environmentalMonitoring}. % need more references
By capturing spatial and spectral data across a wide wavelength range, HSI systems enable material identification, chemical analysis, and anomaly detection with unparalleled precision. Emerging applications in soil nutrient monitoring, food quality assessment, marine ecosystems, and wearable healthcare systems have further underscored the need for portable, low-cost, and high-performance HSI systems \cite{Bacon2004biologyChemistrySpectroscopy,Bae2017wearableHealthcareSpectrometer,Brown2009clinicalUVVisibleSpectroscopy}. On the other hand advanced biomedical applications, such as, disease diagnosis, cancer margin detection, surgical guidance, fetal health monitoring, among other applications, require real-time analysis for efficient decision-making processes \cite{Zhou2023multimodalFLIMOCT,Unger2020realTimeCancerMarginDiagnosis,Smith2020UNMIX_ME}. To keep pace with the demand for these portable HSI systems, an ongoing trend towards the miniaturization of spectrometers is observed, with a focus on achieving chip-scale integration while maintaining high performance (Fig.~\ref{fig:miniaturization_trend}(a)).
Conventional spectrometers rely on dispersive elements such as diffraction gratings or prisms to spatially separate light into its constituent wavelengths. The need to spatially disperse the light makes it challenging to miniaturize, resulting in bulky, delicate, and expensive systems that are unsuitable for portable applications. Advances in microfabrication techniques have enabled the development of bench-top and handheld spectrometers \cite{Correia2000singleChipCMOSopticalSpectrometer,hamamatsu2018microSpectrometer,oceanoptics1992firstminiatureSpectrometer}, but these systems often compromise on performance for a lower footprint. Recent trends show a growing demand for micro and on-chip spectrometers that deliver high performance in a cost-effective and compact form factor \cite{Kulakowski2020marketTrendsSpectrometers}. These on-chip systems are finding increasing applications in biomedical imaging, astronomy, and consumer applications owing to their compact size, low-cost, and ease of integration (Fig.~\ref{fig:miniaturization_trend}(b)) \cite{GoogleScholar}.

\begin{figure}[htbp]
\centering
\includegraphics[width=0.9\textwidth]{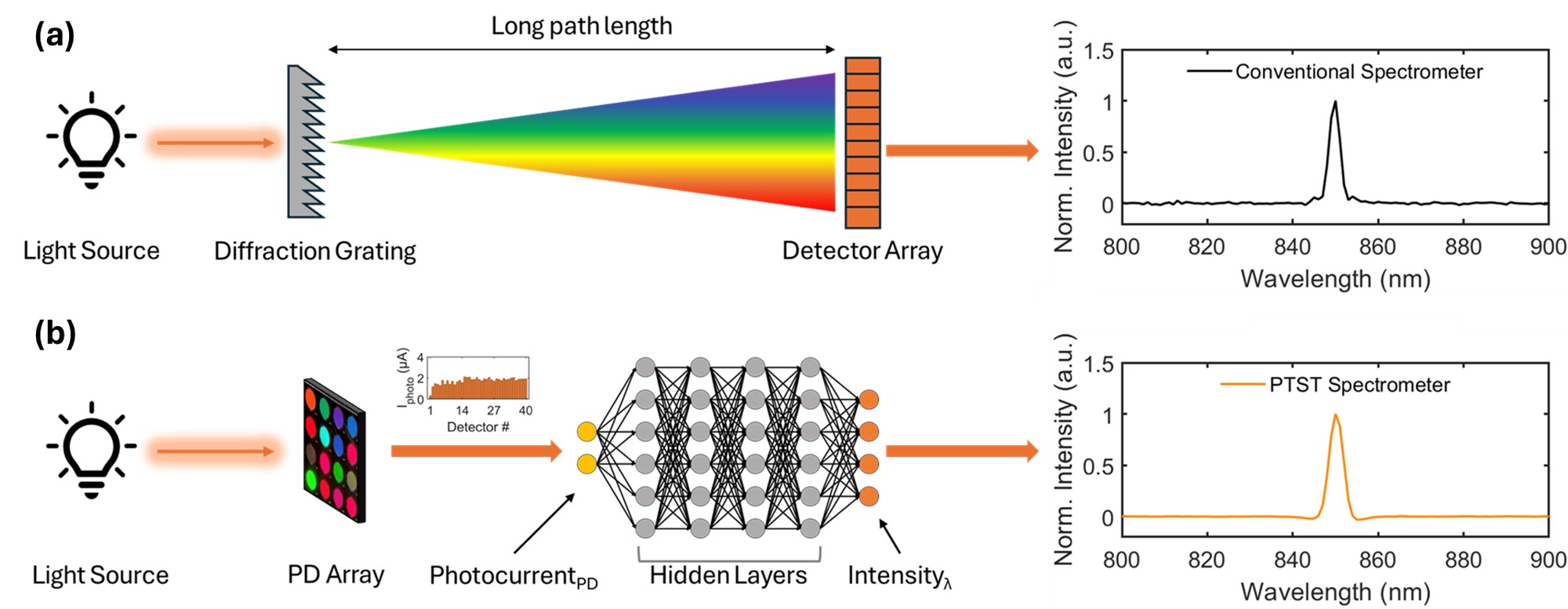}
\caption[Working mechanism of spectrometers]{Working mechanism of spectrometers: (a) Conventional spectrometers with uniform detector arrays disperse the light spatially using diffraction gratings that require long pathlengths owing to their bulky nature. (b) Reconstructive spectrometers utilize unique photodetectors that can capture the minute variations in the incident light spectrum. The spectral information is then reconstructed using machine learning algorithms.}
\label{fig:working_mechanism}
\end{figure}

Recent advances in machine learning and computational power have enabled the development of reconstructive spectrometers that can extract spectral information from a set of unique photodetectors, allowing for miniaturization and cost reduction \cite{yang2021reviewMiniaturizedSpectrometers,Vaswani2017transformer,Xue2024reviewComputationalSpectrometersAlgorithm,Gao2022computationalSpectrometersNanophotonicsandDeepLearning}. 
While light dispersive systems in conventional spectrometers require long path lengths leading to their bulky designs (Fig.~\ref{fig:working_mechanism}(a)), reconstructive spectrometers utilize computational algorithms to extract the spectral information from the captured unique photo-response, enabling compact detector-only spectrometers (Fig.~\ref{fig:working_mechanism}(b)). Researchers have demonstrated unique photodetector arrays using bandgap-engineered semiconductors, meta-surfaces, two-dimensional materials, quantum dots, and other methods \cite{Redding2013compactSpectrometer,Wang2014broabandDiffractiveSpectrometer,Yang2019SingleNanowireSpectrometers,Brown2021plasmonicEncoder,Bao2015colloidalQuantumDotSpectrometer,Wang2019SingleShotSpectralSensors,Huang2017etalonArraySpectrometry,Nitkowski2008microringResonatorSpectroscopy,Yoon2022vanDerWaalsSpectrometers,Zhu2020perovskiteQuantumDotSpectrometer,ahasan2022ReconstructionBasedSpectroscopyRandomPhotonTrappingNanostructure,Zheng2023superconductingNanowireSpectrometer,He2024microSpectrometerTunableOrganicPhotodetector,Faraji-Dana2018metasufaceSpectrometer,Yasunaga2022surfacePlasmonSpectrometer,Zhang2025stressEngineeredSpectrometer}. However, a monolithic on-chip solution is still lacking. %However, the use of such exotic materials makes them costly to fabricate and difficult to integrate into existing systems. 
Silicon-based photodetectors are promising candidates for on-chip integration due to its compatibility with CMOS processes and its well-established fabrication techniques. Some recent works on silicon detectors show promise in achieving spectral diversity including photonic crystal structures \cite{Wang2019SingleShotSpectralSensors}, silicon nanowire photodetectors \cite{Meng2020structurallyColoredSiNanoWires}, and Tapered or Fabry-Perot based silicon detectors \cite{Hu2025siliconNanomembraneSpectrometer,You2024CMOScompatibleFabryPerotSpectrometers}. Some researchers also demonstrated simultaneous spectral and temporal profile extraction using advanced algorithms and ultrafast detectors \cite{Smith2020UNMIX_ME,Ghezzi2021multispectral}. 
% While nanowires and Fabry-Perot based detectors are great for spectral diversification, these structures can be fragile and difficult to fabricate. Tapered detectors on the other hand require fine polishing and are not suitable for large area integration. Photonic crystals can allow transmission engineered spectral response that can be integrated on CMOS image sensors for HSI, however, they also reduce the overall sensitivity of the detector which can be detrimental for low-light applications. Furthermore, these approaches are limited to the visible spectrum and do not extend into the near-infrared (NIR) range. 
One of the significant challenges with silicon detectors, especially in the context of HSI, is their poor performance in the near-infrared (NIR) range. The low absorption coefficient of silicon at longer wavelengths leads to poor responsivity and degraded noise performance, making it difficult to achieve a high signal-to-noise ratio (SNR) in that range. This has limited the applicability of silicon-based spectrometers in the NIR range, where many important spectral features reside \cite{Yao2024chipscaleIRspectrometerforMetrology,cesar2021ApdBiomedicalImaging,Dinish2023miniaturizedVisNIRspectrometerInAgriTech}.

% Update this section to highlight the novelty of this work
In this work, we present a spectrometer-on-a-chip utilizing unique silicon photodiodes (PDs) with sensitivity extending up to \SI{1100}{\nano\meter}. Our spectrometer consists of a set of PDs equipped with unique photon-trapping surface textures (PTST) to achieve spectral diversity across a broad spectral range of \SIrange{640}{1100}{\nano\meter}. We utilize the enhanced light absorption capabilities of PTST to improve the efficiency and sensitivity of the PDs, which in turn boosts the efficacy of hyperspectral imaging of the spectrometer. This spectrometer system is compatible with monolithic integration of readout circuits with conventional CMOS foundry processes. We employ a noise-tolerant fully connected neural network to computationally reconstruct the spectral information from the measured photocurrents. This integration of deep learning represents a key step toward AI-augmented spectral sensing, where neural networks enable compact hardware to achieve high spectral fidelity traditionally possible only with bulky systems. 
This setup allows us to experimentally demonstrate spectral reconstruction of up to $<$ \SI{4}{\nano\meter} narrow-width laser peaks accurately with only 16 unique detectors, achieving \SI{30}{\decibel} higher SNR compared to conventional spectrometers at longer wavelengths. Our noise tolerance analysis demonstrates that the spectrometer can handle additional noise levels of up to \SI{40}{\decibel} from laboratory conditions while maintaining a SNR of \SI{30}{\decibel} even at longer wavelengths. This improved noise performance allows the spectrometer to operate within a broad dynamic range of \SI{50}{\decibel}. Furthermore, our spectrometer-on-a-chip can accurately detect wavelength peaks with \SI{1}{\nano\meter} spectral bins and is capable of resolving two peaks separated at \SI{8}{\nano\meter} apart from each other while operating in the wavelength range of \SIrange{640}{1100}{\nano\meter}. The ability to detect longer wavelengths accurately opens up new possibilities for hyperspectral imaging in biomedical imaging, environmental monitoring, and remote sensing applications. Also, the PD exhibit ultrafast time response of \SI{57}{\pico\second} and high gain ($>$ 7000) making it suitable for high-speed applications like fluorescence lifetime imaging and Raman spectroscopy at low-light conditions. This technology has the potential to revolutionize the field of hyperspectral imaging by providing a compact, cost-effective, and high-performance solution for a wide range of applications.

\section{Results}\label{Results}

\subsection{Spectral Engineering in Silicon Photodiodes}\label{SpectralEngineering}
\begin{figure}[htbp]
\centering
\includegraphics[width=0.7\textwidth]{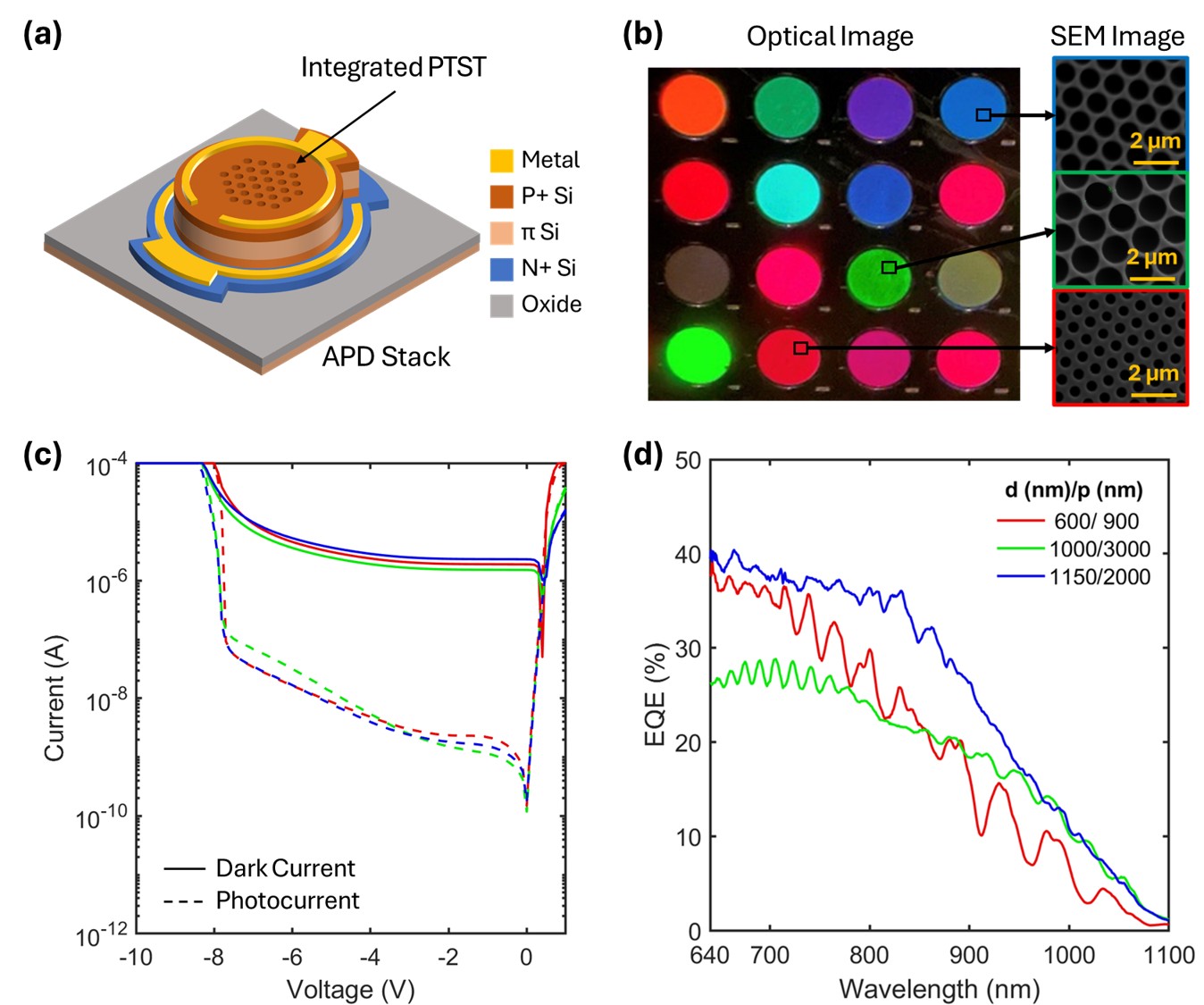}
\caption[Spectral Uniqueness in Photon-trapping photodiode]{Spectral Uniqueness in Photon-trapping photodiode: (a) Schematic drawing of the photodiode (PD) structure with integrated photon-trapping structures (PTST). (b) Optical and SEM images of the PDs. The distinct colorful reflection from the PDs is due to the presence of different PTSTs designed to be optically unique. (c) IV profile of selected PDs under dark and illuminated conditions with (\SI{10}{\micro\watt} laser power at \SI{800}{\nano\meter} wavelength). (d) Measured external quantum efficiency of selected PDs across the \SIrange{640}{1100}{\nano\meter} wavelength range demonstrating improved responsivity at different wavelengths due to PTST enhancement.}
\label{fig:unique_detectors}
\end{figure}

Silicon photodiodes (PDs) are widely used in visible wavelength applications due to their high sensitivity and low noise characteristics. However, their poor performance in the NIR range is limiting their applicability. Recent works on PTST have shown promise in enhancing the light absorption capabilities of silicon PDs, specifically in the NIR wavelength range \cite{Gao2017highSpeedSiPhotodiodeNIRwavelength,ahasan2020SmartNanophotonics,wayesh2023AchievingHigherPhotoabsorptionThanIII-V}. PTST can interact with incident light to trap photons in lateral propagating modes, thereby enhancing the effective path length and improving absorption and responsivity \cite{Gao2017photonTrappingMicrostructures}.
Furthermore, previous works have demonstrated that modulating the dimensions of PTST can lead to controlled photon penetration depth and strong coupling for specific wavelengths of light, enabling enhanced external quantum efficiency (EQE) at those wavelengths \cite{ahamed2022optimizingLightPenetration,ahasan2023UniqueHyperspectralResponse}. Utilizing this property, we can engineer the spectral response of the silicon PDs by designing PTSTs with varying dimensions. This leads to distinctive spectral responses in each PDs. The simulated absorption spectra for different PTST designs can be found in the Supplementary Information Sec. \ref{supp-PTST_design}. 
The avalanche photodiode (APD) stack is designed with a ~P$^+$~--~$\pi$~--~P~--~N$^+$ doping profile to achieve low breakdown voltage and high gain. The PTSTs are integrated into the APDs to engineer the photoresponsivity of the silicon PDs in a wide spectral range, as shown in Fig.~\ref{fig:unique_detectors}(a). Details of the fabrication process and device structure are provided in Sec. \ref{Methods} and Supplementary Information Sec. \ref{supp-Device_Fabrication}.
The unique photoresponsivity of the APDs with PTST is evidently captured under the white light of an optical microscope in the form of iridescence, and their structural differences are shown in the SEM image (Fig.~\ref{fig:unique_detectors}(b)). Here each PTST is reflecting certain wavelengths out of the entire white light spectra based on the optical bandgap formation due to variation in their structures \cite{ahasan2023UniqueHyperspectralResponse}. The IV measurement of the three unique PDs under dark and illuminated conditions is shown in Fig.~\ref{fig:unique_detectors}(c). The devices were illuminated with a laser source centered at \SI{800}{\nano\meter} with an optical power of \SI{10}{\micro\watt} incident on the device. The difference in the photocurrent in the PDs under the same illumination is clearly visible from the IV profile. The devices also show a low breakdown voltage of \SI{7.8}{\volt} and a low dark current of the order of \SI{}{\nano\ampere} at the onset of breakdown. The EQE of the respective PDs is measured at a unity gain voltage of \SI{2}{\volt} across the wavelength range of \SIrange{640}{1100}{\nano\meter} and is shown in Fig.~\ref{fig:unique_detectors}(d). The EQE profiles of the three PDs are distinctive from each other due to their unique PTST design. These devices are optimized for approximately \SI{600}{\nano\meter}, \SI{800}{\nano\meter}, and \SI{1000}{\nano\meter} wavelengths represented as the red, blue, and green EQE curves respectively. The measured EQE profiles of all the fabricated PDs are provided in the Supplementary Information Sec. \ref{supp-EQE_PTST}. This CMOS compatible unique PTST designs allow us to achieve spectrally engineered PDs monolithically integrated with the readout circuitry in a conventional CMOS foundry.
    
%The EQE of the PTST enhanced PDs are measured to be as high as 0.33 at \SI{850}{\nano\meter} wavelength, whereas a flat PD without PTST shows EQE of only 0.05 at the same wavelength. 

\subsection{Model Training}\label{ModelTraining}
\begin{figure}[htbp]
\centering
\includegraphics[width=0.9\textwidth]{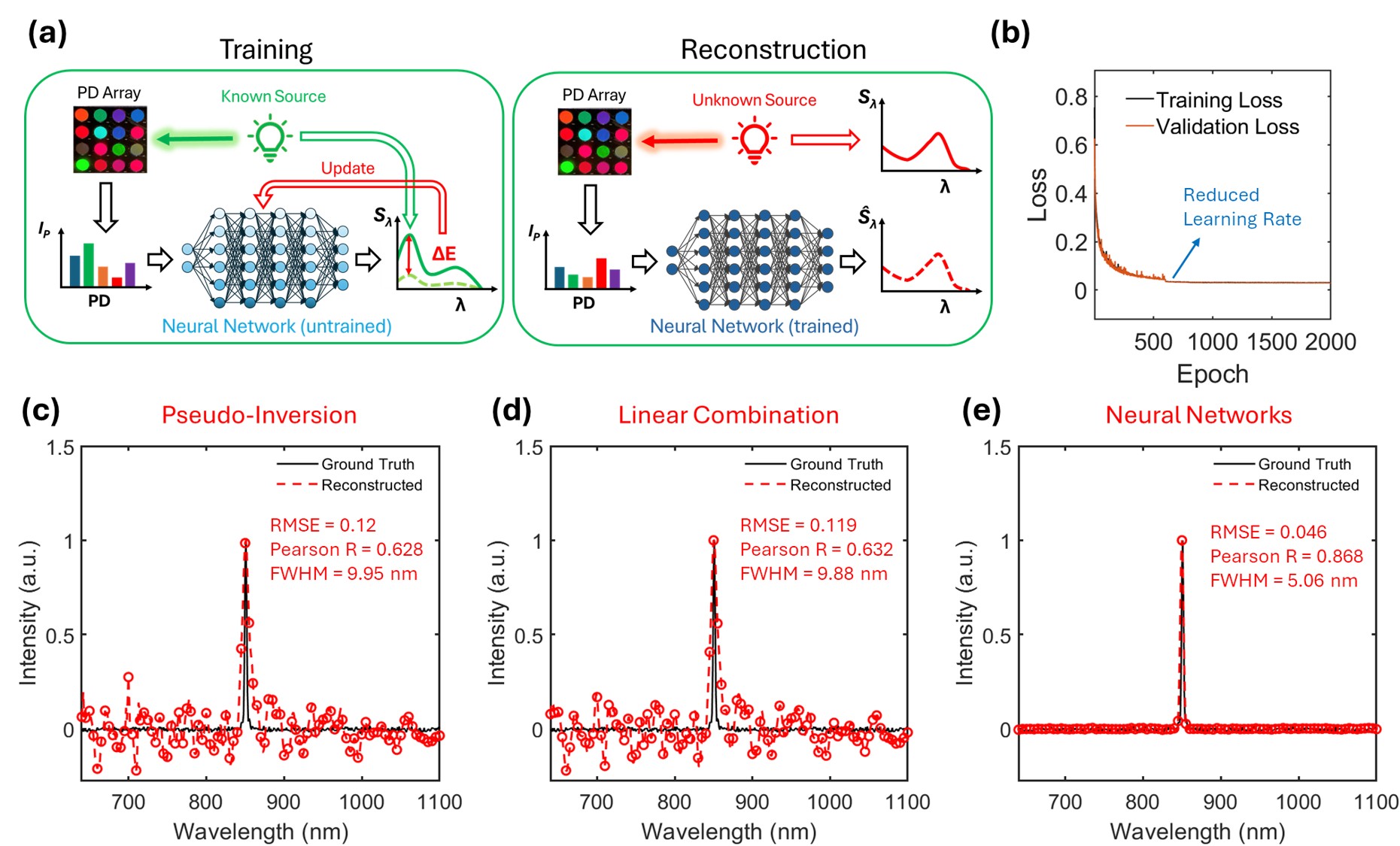}
\caption[Neural network model for spectral reconstruction]{Neural network model for spectral reconstruction: (a) Demonstration of the training and reconstruction process of the neural network. (b) Training and validation loss plotted against epoch showing convergence of the model. The model is trained for 2000 epochs with the loss function converging around 0.03. (c-e) Comparison of spectral reconstruction using (c) matrix pseudo-inversion, (d) linear combination of Gaussian functions, and (e) neural network model. The neural network model outperforms the other two methods in reconstructing the spectral profile of \SI{3}{\nano\meter} FWHM laser peak. The RMSE and Pearson's R value for the neural network model are 0.046 and 0.87 respectively indicating high accuracy in spectral reconstruction.
}
\label{fig:neural_network}
\end{figure}

The spectral reconstruction of the PTST spectrometer is performed using a fully connected neural network. The neural network architecture consists of an input layer, four hidden layers, and an output layer. The input layer receives the measured photocurrents from the PDs, while the output layer provides the reconstructed spectral information. The hidden layers consist of fully connected ``neurons" that are trained to map the input to the output over several iterations. The neural network is trained using a synthetic dataset of more than 500,000 spectra and their corresponding expected photocurrents, obtained by applying the measured EQE of the PDs. The dataset is generated using a combination of Gaussian functions with varying peak wavelengths, full width at half maximum (FWHM), and amplitudes to simulate the spectral profile of the light source.
Figure~\ref{fig:neural_network}(a) demonstrates the training and reconstruction process of the neural network. The training process involves feeding the neural network with the photocurrents from the PDs and their corresponding spectral information. The neural network then calculates the loss function from its estimated spectra and the true value. By back-propagating the loss function through each layer, the model updates its weights and biases at each neuron. This way the network ``learns" the relationship between the spectral profile and the corresponding photocurrents. We used a custom loss function using root mean squared error (RMSE) and Pearson's correlation coefficient (R) that gave us the best results. The model architecture and the loss functions are discussed in detail in the Sec. \ref{Methods} and Supplementary Information Sec. \ref{supp-Neural_Network_Model}. The training and validation loss values for the model training process are plotted against each epoch in Fig.~\ref{fig:neural_network}(b). The training and validation losses decrease rapidly below 0.1 after 100 epochs and converge to around 0.03 after 1000 epochs. This indicates that the model is learning well and can generalize to unseen data. The learning rate was reduced by a factor of 10 after 600 epoch for better convergence. The model is then used to reconstruct the spectral information from the measured photocurrents of the PDs. The reconstructed spectral profile is compared with other reconstruction methods in Fig.~\ref{fig:neural_network}(c-e). Matrix pseudo-inversion is the simplest method used for spectral reconstruction, but it suffers from noise sensitivity and does not perform well for sharp spectral signals as observed in Fig.~\ref{fig:neural_network}(c) \cite{moore1920reciprocal}. Linear combination of Gaussian or sinusoidal functions is another approach that works well for broad spectra \cite{Donoho2006CompressedSensing,ahasan2022ReconstructionBasedSpectroscopyRandomPhotonTrappingNanostructure,Wang2019SingleShotSpectralSensors,ahasan2024OnChipHyperspectralDetectors}. However, their performance for laser peaks with narrow spectral width is similar to that of matrix pseudo-inversion (Fig.~\ref{fig:neural_network}(d)). The RMSE and Pearson's R for both cases are $\sim$0.12 and $\sim$0.63, respectively. Neural networks, on the other hand, can learn the complex relationship between the spectral profile and the photocurrents of the PDs, allowing them to accurately predict the sharp laser spectra with high accuracy, achieving RMSE of 0.046 and Pearson's R value of 0.87 (Fig.~\ref{fig:neural_network}(e)). Furthermore, their inherent noise tolerance allows them to perform well even in the presence of noise. A detailed discussion of the noise tolerance analysis is provided in Sec. \ref{NoiseTolerance}.
    %The neural network model is able to reconstruct the spectral profile with a RMSE of 0.03 and a Pearson R value of 0.9 for the laser peaks. This indicates that the neural network model is able to accurately reconstruct the spectral information from the measured photocurrents.

\subsection{Spectral Reconstruction}\label{SpectralReconstruction}
For the unknown light source, we used a supercontinuum laser (NKT SuperK Extreme EXR-04) coupled with an opto-acoustic filter (NKT Select NIR) with narrow spectral bandwidth of $<$\SI{4}{\nano\meter} to evaluate the performance of the PTST spectrometer. The filter operates at a range of \SIrange{640}{1100}{\nano\meter}, thus the characterization of the spectrometer is performed within this range. For each laser spectrum, the measured photocurrents from the spectrally unique PDs are fed into the trained neural network for spectral reconstruction. For comparison, a conventional silicon spectrometer is used as the ground truth for spectral reconstruction.

\begin{sidewaysfigure}
\centering
\includegraphics[width=0.9\textwidth]{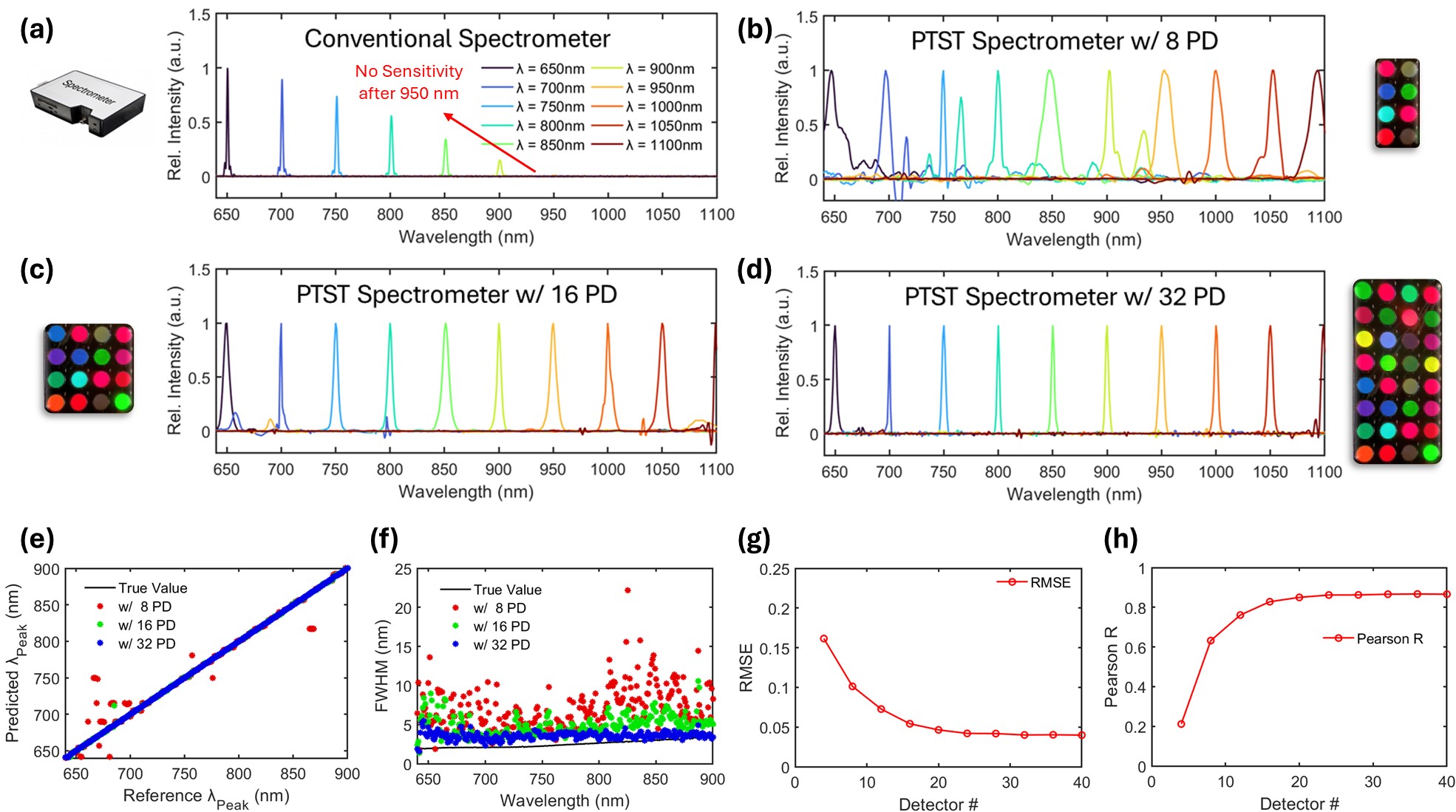}
\caption[Spectral reconstruction of laser peaks]{Spectral reconstruction of laser peaks: (a-d) Measured spectral profile of narrow band laser peaks (\SI{10}{\micro\watt} optical power) (a) in a conventional silicon spectrometer compared with reconstructed spectral profile in the PTST spectrometer using measured photocurrents from (b) 8, (c) 16, and (d) 32 photodiodes (PDs) respectively for wavelengths ranging from \SIrange{640}{1100}{\nano\meter}. 
The conventional spectrometer shows no sensitivity beyond 950 nm while the PTST spectrometers can detect the laser peaks accurately up to 1100 nm. The reconstructed spectra with 8 PDs show poor accuracy due to lack of spectral information. However, the reconstruction quality improves markedly as the number of PDs increases. All spectral intensities are normalized to the maximum count across the measured spectra. 
(e) The peak wavelength accuracy results indicate that the PTST spectrometer performs with high precision when using 16 and 32 PDs, accurately predicting the peak wavelengths. However, while only 8 PDs, the prediction accuracy declines noticeably. (f) Comparison of full width at half maximum (FWHM) of the PTST spectrometers shows that it can almost resemble the narrow width of the spectral profile achieving $<$\SI{4}{\nano\meter} FWHM for 16 and 32 PDs. (g-h) The reconstruction accuracy of the PTST spectrometer in terms of (g) average root mean squared error (RMSE) and (h) Pearson's correlation coefficient (R) are plotted against the number of detectors used for reconstruction of incident laser peaks in the range of \SIrange{640}{900}{\nano\meter}. The average RMSE decreases to $<$0.05 while the Pearson's R value increases to $>$0.85 with 32 detectors, demonstrating a good fit with the ground truth. 
% Increasing the number of detector further shows no improvement in reconstruction accuracy indicating that no new information is added from the other PDs due to inadequate spectral encoding.
}
\label{fig:laser_reconstruction}
\end{sidewaysfigure}

Figure~\ref{fig:laser_reconstruction} shows the comparison between the spectral reconstruction of narrow band laser peaks in the PTST spectrometer with 8, 16 and 32 PDs against the measured spectral profile in a conventional silicon spectrometer for wavelengths ranging from \SIrange{640}{1100}{\nano\meter}. The optical power of the narrow band laser peak is calibrated to be \SI{10}{\micro\watt}. 
In the conventional spectrometer, the measured photon counts are translated into relative intensity, maintaining the same level of background counts. To do that, the intensity is normalized using the maximum count among all measured spectra. We observed that beyond \SI{900}{\nano\meter}, the conventional spectrometer shows poor sensitivity and beyond \SI{950}{\nano\meter}, the background noise begins to dominate. This is caused by the poor absorption capability of Si in the NIR wavelengths, contributing to the SNR reduction resulting in nontraceable signals beyond \SI{950}{\nano\meter}. 
On the other hand, the PTST spectrometers with 8, 16 and 32 PDs can detect the narrow band laser peaks with high accuracy and SNR up to \SI{1100}{\nano\meter}. The reconstructed spectra are similarly normalized to relative intensity by dividing each spectrum with the maximum intensity among all reconstructed spectra. Here, we observed improved sensitivity at longer wavelengths due to the integration of PTST, which plays a crucial role in trapping light. Furthermore, the PTST spectrometer relies on the unique responsivity of the PDs rather than the relative count rates, which allows it to detect longer wavelengths more accurately compared to conventional silicon spectrometers. The Supplementary Information Sec. \ref{supp-Unique_Response} contains the measured photocurrent from the PDs for the laser peaks in the range of \SIrange{640}{1100}{\nano\meter} and further justification for the improved sensitivity of the PTST spectrometer at longer wavelengths.

Comparing the reconstructed spectra with the ground truth data from the conventional spectrometer in the range of \SIrange{640}{900}{\nano\meter}, we observe that the PTST spectrometer with 8 PDs shows poor peak accuracy due to lack of spectral information. However, for a higher number of detectors, the PTST spectrometer can easily detect the peak wavelengths (Fig.~\ref{fig:laser_reconstruction}(e)). 
Figure~\ref{fig:laser_reconstruction}(f) shows the comparison of the FWHM of the PTST spectrometers with the ground truth data from the conventional spectrometer. The FWHM of ground truth laser peaks vary from \SIrange{2.5}{4}{\nano\meter} with increasing wavelengths (black solid line). The PTST spectrometer can predict the FWHM closely with the best case for 32 PDs showing a FWHM of $<$\SI{4}{\nano\meter}. 
The average RMSE and Pearson's R are plotted against the number of detectors used for incident laser peaks in Fig.~\ref{fig:laser_reconstruction}(g-h). The average RMSE is high for lower number of detectors, but it decreases to $<$0.05 with 16 PDs and saturates to 0.04 with 32 PDs. Increasing the number of detectors further shows no improvement in reconstruction accuracy indicating that no new information is added from the other PDs due to inadequate spectral diversity. The contrast in Pearson's R value is much higher, with an improvement from 0.2 to 0.85 for the PTST spectrometer with 4 to 32 PDs, respectively. Pearson's R value is a better indicator of reconstruction accuracy for sharp spectral features and, therefore, is used in the loss function along with RMSE. PTST spectrometer with only 16 PDs can achieve an RMSE of $<$0.05 and Pearson's R value of $>$0.8, indicating high spectral encoding capability of the PTST with a limited number of detectors.

\subsection{Noise Tolerance}\label{NoiseTolerance}

\begin{figure}[htbp]
\centering
\includegraphics[width=1\textwidth]{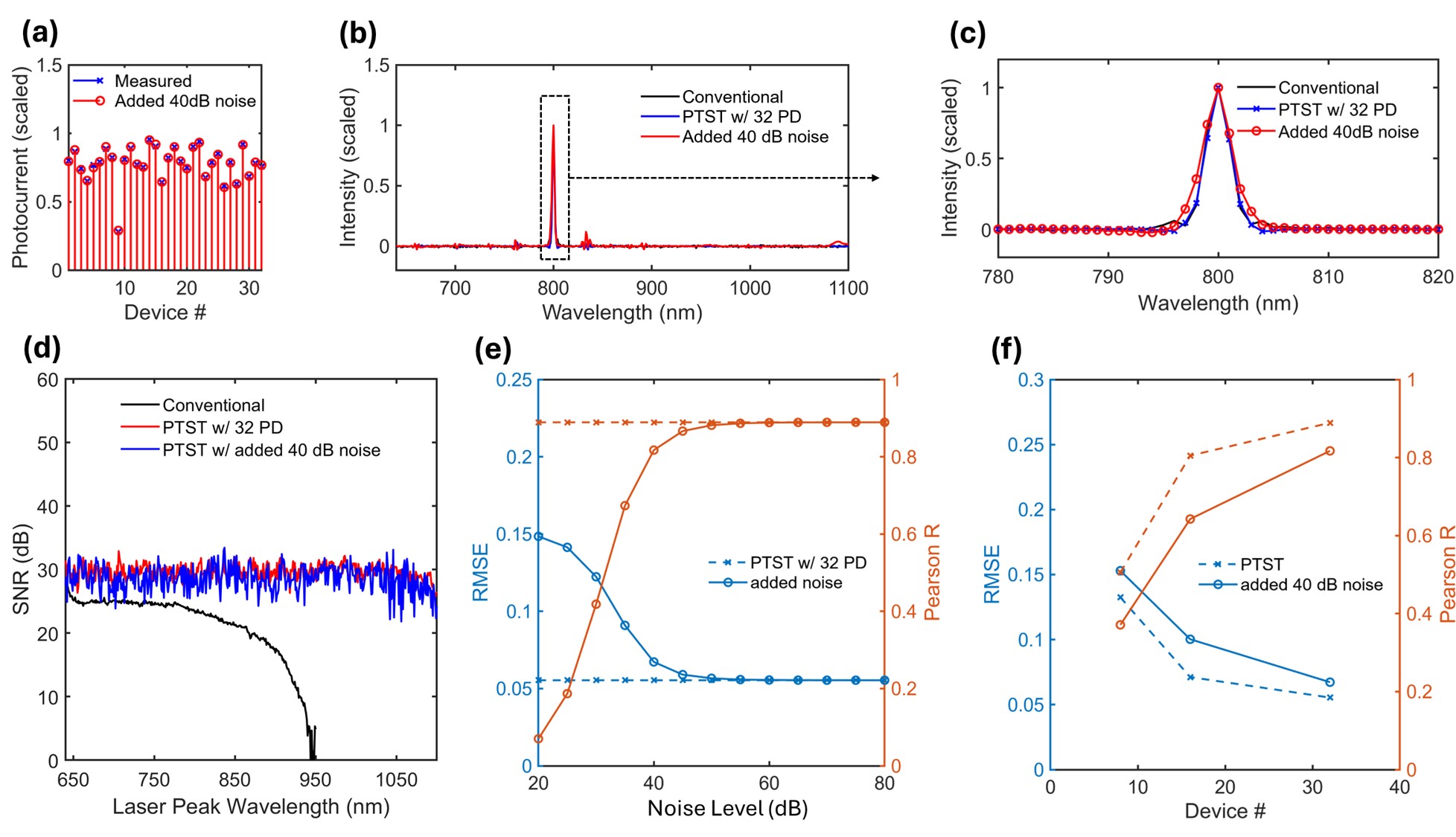}
\caption[Noise tolerance in PTST spectrometer]{Noise tolerance in PTST spectrometer: (a) The measured photocurrent for 32 PDs against their respective PDs are plotted in blue crosses which includes all experimental noise. Simulated white noise of \SI{40}{\decibel} signal-to-noise ratio (SNR) is added to the measured photocurrents depicted in red circles. (b) reconstructed spectral profile from the PTST spectrometer is compared for its base performance and added noise case against the ground truth spectra from the spectrometer (intensity is scaled to unity). (c) Zoomed in view of the spectral profile depicting slight broadening of the spectra due to added noise. (d) The SNR of the reconstructed spectrum for the base and noise added case is compared with the measured spectrum from a conventional silicon spectrometer. PTST spectrometer demonstrates \SI{30}{\decibel} SNR in laboratory conditions and can maintain same level of sensitivity with additional \SI{40}{\decibel} detector noise. Whereas the conventional spectrometer can only maintain \SI{25}{\decibel} SNR up to \SI{800}{\nano\meter} after which the SNR drops quickly to \SI{0}{\decibel} beyond \SI{950}{\nano\meter}. (e) The combined RMSE, and Pearson R value is plotted on the left and right respectively against added detector noise levels for laser peaks ranging from \SIrange{640}{900}{\nano\meter}. The PTST spectrometer can perform significantly well for a noise level up to \SI{40}{\decibel} SNR, afterwards, the reconstruction accuracy drops drastically as depicted in the Pearson R value. (f) The combined RMSE and Pearson R value is measured for a noise level of \SI{40}{\decibel} SNR in PTST spectrometer with 8, 16, and 32 PDs, respectively. Reduction of devices leads to lower noise tolerance and lower reconstruction accuracy.}
\label{fig:noise_tolerance}
\end{figure}

Noise is a critical factor in the performance of spectrometers, especially in low-light applications. Since the PTST spectrometer relies on the photocurrent measurement from the PDs, any noise in the PD can significantly affect the spectral reconstruction accuracy. Therefore, we analyzed the noise tolerance of the PTST spectrometer under ambient conditions and with added white noise to the detector current; the results are also compared with a conventional silicon spectrometer. The measured SNR of the PDs under \SI{10}{\micro\watt} illumination with a peak wavelength of \SI{800}{\nano\meter} is of the order of \SI{60}{\decibel}. This considers all background and intrinsic PD noise in the experimental measurement setup. The measured photocurrents for 32 PDs are shown in Fig.~\ref{fig:noise_tolerance}(a) as blue crosses. To explore the performance of the spectrometer in a noisier environment, we added simulated white noise equivalent to a \SI{40}{\decibel} SNR to the measured photocurrents. The noise added photocurrents are shown as red circles in Fig.~\ref{fig:noise_tolerance}(a). 

Now both sets of photocurrents are fed into the neural network for spectral reconstruction. For \SI{800}{\nano\meter} peak wavelength, the reconstructed spectral profile from the PTST spectrometer is compared for its base performance and added noise case against the ground truth spectra from the conventional spectrometer (intensity is scaled to unity) in Fig.~\ref{fig:noise_tolerance}(b-c). The zoomed-in view of the spectral profile in Fig.~\ref{fig:noise_tolerance}(c) depicts slight broadening of the spectra due to the added \SI{40}{\decibel} noise, but the spectra are resolved accurately.
The SNR of the reconstructed spectra is measured to be $\sim$\SI{30}{\decibel} for the experimental setup across the spectral range of \SIrange{640}{1100}{\nano\meter} as shown in Fig.~\ref{fig:noise_tolerance}(d). With the additional \SI{40}{\decibel} detector noise in the photocurrent, the PTST spectrometer can still maintain $\sim$\SI{30}{\decibel} SNR with a slight drop in performance. Whereas the conventional spectrometer demonstrates $\sim$\SI{25}{\decibel} SNR up to \SI{800}{\nano\meter} and quickly drops to \SI{0}{\decibel} beyond \SI{950}{\nano\meter}. This indicates that the PTST spectrometer can tolerate significantly higher noise levels than conventional spectrometers while maintaining its performance. This can be attributed to the higher sensitivity of the PTST enhanced PDs and clean training dataset. Note that the SNR drops slightly after \SI{1050}{\nano\meter} wavelength due to the intrinsically poor efficiency of silicon PDs.
Similar test with varying noise levels was studied and the combined RMSE and Pearson R value is plotted on the left and right Y-axes, respectively against increasing SNR for laser peaks ranging from \SIrange{640}{900}{\nano\meter} in Fig.~\ref{fig:noise_tolerance}(e). The PTST spectrometer can perform significantly well for added detector noise level up to \SI{40}{\decibel} SNR, afterwards, the reconstruction accuracy drops drastically as depicted in the RMSE and Pearson R value. The combined RMSE and Pearson R value is measured for a noise level of \SI{40}{\decibel} SNR in the PTST spectrometer with 8, 16, and 32 PDs, respectively, and is shown in Fig.~\ref{fig:noise_tolerance}(e). As expected, the reduction of the number of devices leads to lower noise tolerance and lower reconstruction accuracy.

\subsection{Spectrometer Performance}

\begin{figure}[htbp]
\centering
\includegraphics[width=0.9\textwidth]{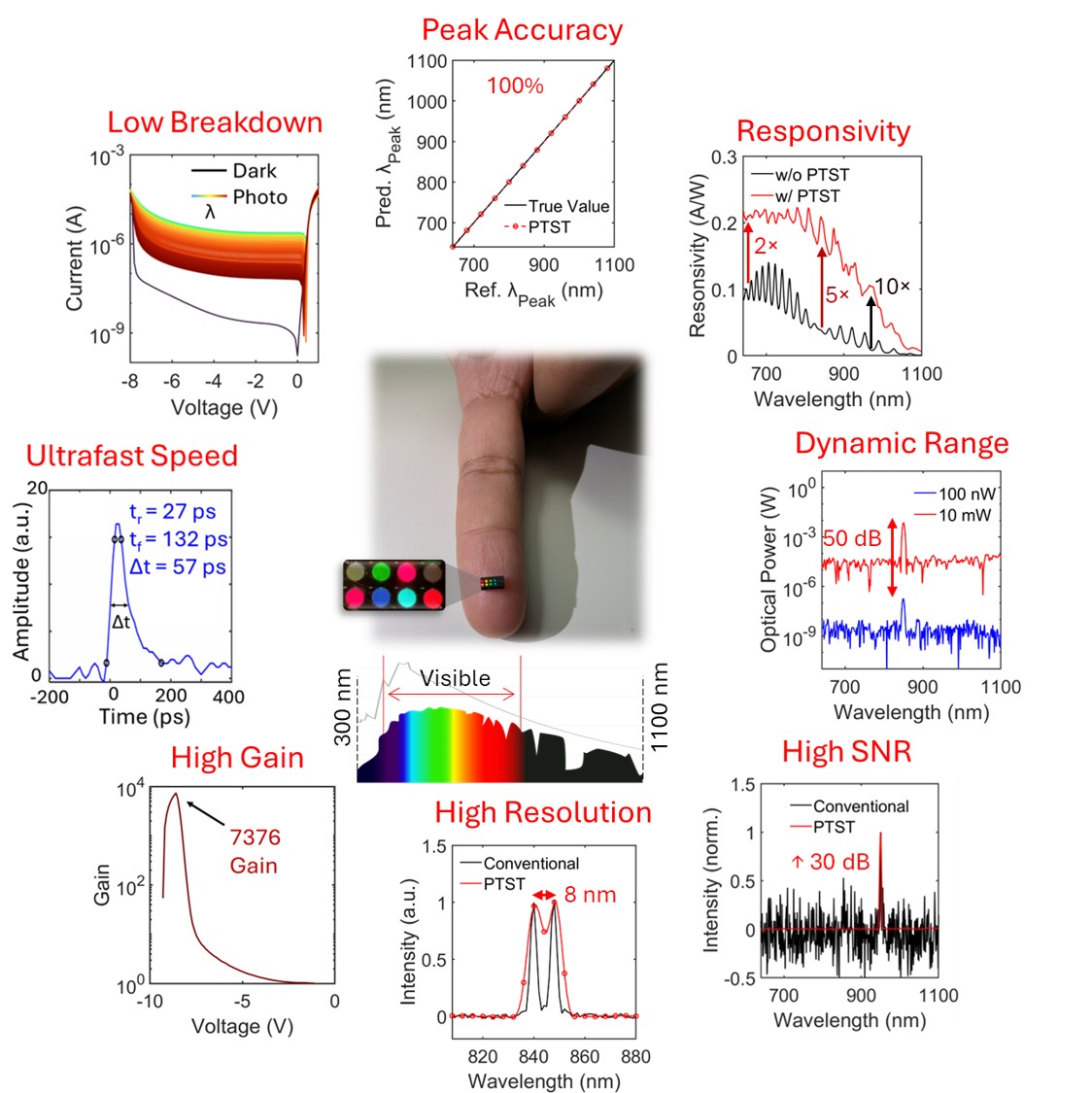}
\caption[Performance metric of spectrometer-on-a-chip]{Performance metric of spectrometer-on-a-chip: The developed spectrometer-on-a-chip is placed on a fingertip for size comparison. The colorful patterns on the chip represent the photon-trapping surface textures (PTST) that are responsible for the unique spectral response of the PDs. Even with this small footprint, the spectrometer exhibits up to 10$\times$ higher responsivity, a dynamic range of \SI{50}{\decibel}, and a high SNR of \SI{30}{\decibel}.
The inherent photoresponsivity of silicon PDs in the wavelength range of \SIrange{300}{1100}{\nano\meter} allows the spectrometer to operate in a wide spectral range.
The integrated spectrometer has been experimentally demonstrated in the wide spectral range of \SIrange{640}{1100}{\nano\meter} with spectral resolution of \SI{8}{\nano\meter} demonstrating $100\%$ peak accuracy. Furthermore, the ultrafast time response of \SI{57}{\pico\second}, combined with \SI{7.8}{V} breakdown voltage and high gain ($>$7000) of the PDs allows it to perform well in photon-starved conditions. 
}
\label{fig:spectrometer_on_a_finger_tip}
\end{figure}

The developed spectrometer-on-a-chip is shown in Fig.~\ref{fig:spectrometer_on_a_finger_tip} placed on a fingertip for size comparison. The unique PTST design integrated onto the PDs enable spectrally unique responsivity across multiple PDs, as observed from the reflected colors from the chip. Here larger area PDs (\SI{500}{\micro\meter} diameter) are shown for better visual. The spectrometer can have a small footprint of \SI{0.4}{\milli\meter\squared} (using \SI{100}{\micro\meter} diameter PDs) and is responsive over a wide spectral range of \SIrange{300}{1100}{\nano\meter} due to silicon's inherent photoresponsivity. 
The PTST improves the responsivity of the PDs significantly up to 10$\times$ at $\sim$ \SI{950}{\nano\meter} wavelength compared to a PD without PTST, thereby enabling better sensitivity and dynamic range. 
The spectral resolution of the spectrometer is measured to be \SI{8}{\nano\meter} with $100\%$ peak accuracy and $\sim$0.05 spectral reconstruction error (in RMSE) while operating in the spectral range of \SIrange{640}{1100}{\nano\meter}. Note that the spectral resolution is limited by the spectral width of the laser peaks used for characterization, which is $\sim$\SI{4}{\nano\meter} in this case. The spectrometer-on-a-chip is demonstrated to have a dynamic range of \SI{50}{\decibel} and a SNR of \SI{30}{\decibel} making it suitable for on field applications. Details on the spectral resolution and dynamic range measurements are provided in the Supplementary Information Sec. \ref{supp-Spectral_Resolution} and \ref{supp-Dynamic_Range}. 
The ultrafast time response of the PDs is measured to be \SI{57}{\pico\second} making it suitable for high-speed applications like fluorescence lifetime imaging and Raman spectroscopy. The low breakdown voltage of \SI{7.8}{\volt} paired with high gain ($>$ 7000) allows the spectrometer to operate at low power levels and provide high sensitivity even in low light conditions. 
% The spectrometer-on-a-chip is a promising solution for miniaturized, low-cost, and high-performance hyperspectral imaging applications.

    %% Need to add more lines and limit redundancy and repetition %%

\subsection{Hyperspectral Imaging}\label{HyperspectralImaging}

\begin{figure}[htbp]
\centering
\includegraphics[width=1\textwidth]{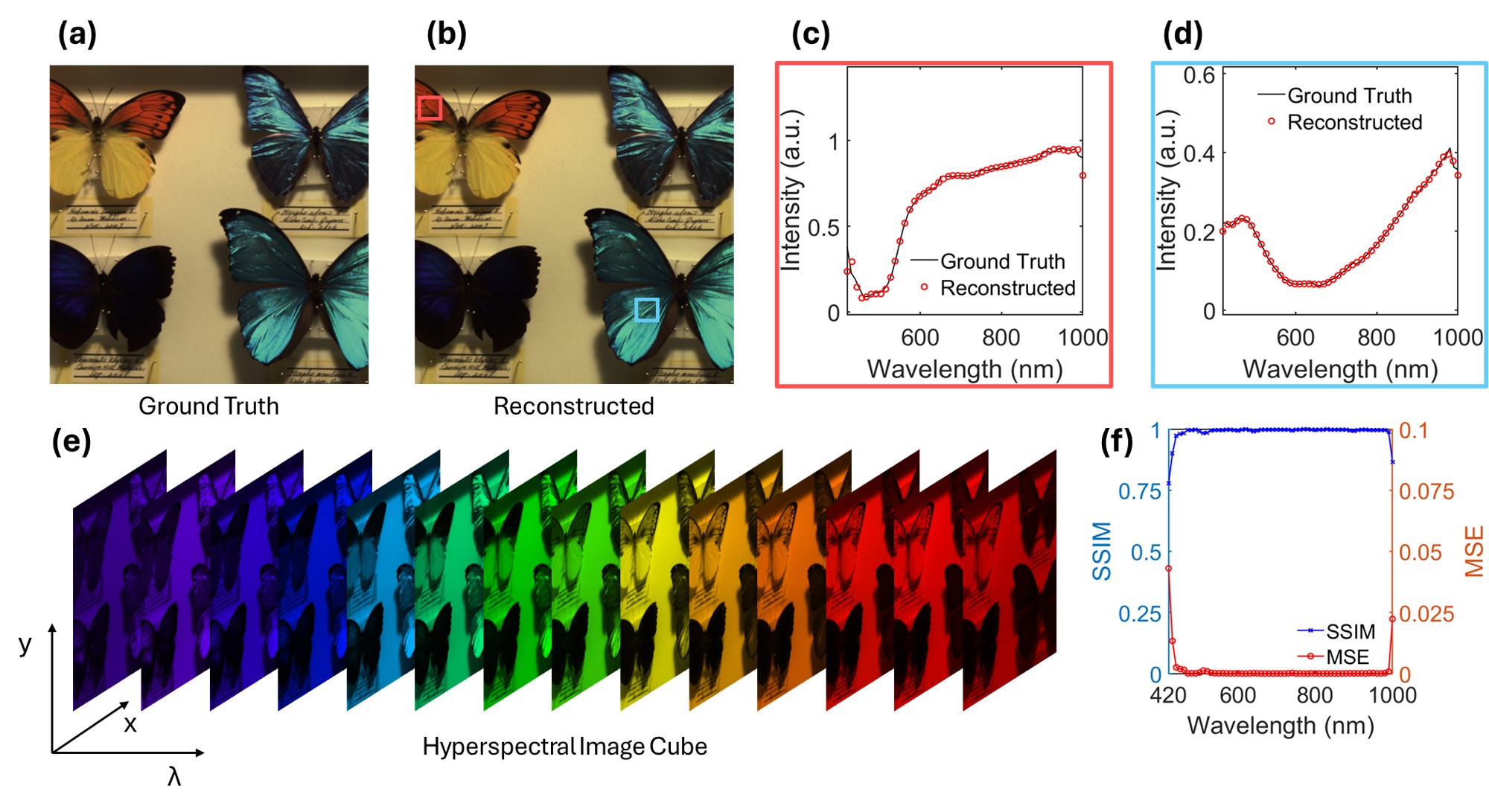}
\caption[Hyperspectral Image of Butterflies]{Hyperspectral Image of Butterflies: (a) The ground truth image of the Butterflies dataset \cite{Yusuke2019Butterfly}. (b) The reconstructed hyperspectral image using simulated spectral response of 30 PDs. (c-d) The reconstructed hyperspectral spectrum for (c) pixel (10,100) and (d) pixel (400,400) highlighted in red and blue boxes in (b). The reconstructed spectra closely match the ground truth spectra, depicting the red and blue color pigments accurately in the Butterflies. (e) The reconstructed hyperspectral images at different wavelengths. (f) The spectral reconstruction accuracy of the spectrometer-on-a-chip for the Butterflies dataset over the spectral bands. The mean squared error (MSE) is close to zero over the broad spectral range and structural similarity index metrics (SSIM) is close to unity with an average MSE of 0.0015 and an average SSIM of 0.9875. There is a slight increase in error near the edge of the wavelength range which can be attributed to the poor EQE diversity of the PDs at those range. Overall, the results demonstrate the high accuracy and fidelity of the AI-enabled spectrometer-on-a-chip for hyperspectral imaging applications.}
\label{fig:HSI_butterfly}
\end{figure}

To demonstrate the capabilities of the spectrometer-on-a-chip for hyperspectral imaging (HSI), we used an open-source hyperspectral dataset of a butterflies \cite{Yusuke2019Butterfly}. The dataset contains 59 spectral bands ranging from \SIrange{420}{1000}{\nano\meter} with \SI{10}{\nano\meter} separations. The dataset is available in the form of a 3D hyperspectral image cube with a spatial resolution of $512 \times 512$ pixels. For this proof-of-concept demonstration, we used simulated spectral response of the PDs as the EQE of the PDs in the shorter wavelengths are not available. The simulated spectral response of 30 PDs is generated by simulating the optical absorption of light in the silicon PDs with different PTST using finite distance time domain (FDTD) method with the help of Lumerical FDTD Solutions Tool. The simulated responsivity can be found in the Supplementary Information Fig. \ref{supp-supfig:simulated_eqe}. 
The hyperspectral dataset was interpolated to 581 bins with \SI{1}{\nano\meter} separation in the range from \SIrange{420}{1100}{\nano\meter}. Then it was integrated with the simulated spectral response of the PDs to generate their photocurrents which was later used to reconstruct the hyperspectral image using the neural network model. 
The ground truth image of the Butterflies dataset and the reconstructed hyperspectral image are shown in Fig.~\ref{fig:HSI_butterfly}(a) and (b) respectively. The reconstructed hyperspectral spectrum for pixel (10,100) and pixel (400,400) highlighted in red and blue boxes in Fig.~\ref{fig:HSI_butterfly}(b) are compared with the ground truth in Fig.~\ref{fig:HSI_butterfly}(c-d), respectively. The reconstructed spectra closely match the ground truth spectra capturing the minute details accurately. Figure~\ref{fig:HSI_butterfly}(e) shows the reconstructed hyperspectral image cube at different wavelengths, while Fig.~\ref{fig:HSI_butterfly}(f) presents the spectral reconstruction accuracy of the spectrometer-on-a-chip for the Butterflies dataset over the spectral bands, evaluated using the mean squared error (MSE) and structural similarity index metrics (SSIM) of the reconstructed image compared with the ground truth. The average MSE and SSIM across all reconstructed wavelengths is 0.0015 and 0.9875 respectively. The increased MSE near the shorter wavelength edge can stem from the lack of unique spectral response of the PDs, while poor sensitivity is more responsible for the discrepancy at the longer wavelength edge. However, the overall results demonstrate a high degree of accuracy and fidelity for the spectrometer system. Furthermore, the use of AI-augmented spectral reconstruction enables the system to extract accurate hyperspectral data even with limited hardware, showcasing its potential for real-world imaging tasks.

\begin{sidewaystable}[htbp]
\centering
\caption{Benchmarking of Computational Reconstructive Spectrometers}
\begin{tabular}{|p{4cm}|p{0.75cm}|p{1.3cm}|p{0.7cm}|p{1.1cm}|p{2.35cm}|p{1.4cm}|p{1.3cm}|p{1.6cm}|p{1.2cm}|}
\hline
\textbf{Material / Sensor} & \textbf{Year} & \textbf{Area} & \textbf{Det. \#} & \textbf{CMOS Comp.} & \textbf{Spectral Bandwidth} & \textbf{Spectral Res.} & \textbf{Timing Profile} & \textbf{SNR (Det. Noise)} & \textbf{Illu.} \\
\hline
Electrochromic filter on CMOS sensor \cite{Tian2024electrochromicSpectrometer} & 2024 & 1 \SI{}{\centi\meter\squared} & 1 & Yes & 400--800 nm & $<$10 nm & NA & ($>$37 dB) & Surface \\
\hline
InSe/NbTe$_2$ 2D tunable photodetector \cite{Cui2025tunableOptoelectronicInterface} & 2025 & 5$\times$5 \SI{}{\micro\meter\squared} & 1 & No & 500--840 nm & 0.19--2.45 nm & Yes & NA & Surface/ Waveguide \\
\hline
Etched TiO$_2$ metasurface on CMOS sensor \cite{Tang2024metasurfaceSpectrometers} & 2024 & 1 \SI{}{\milli\meter\squared} & 100 & Yes & 480--610 nm & 1.7--1.9 nm & NA & $\sim$25 dB & Surface \\
\hline
CdS$_x$Se$_{1-x}$ nanowire array \cite{Yang2019SingleNanowireSpectrometers} & 2019 & $\sim$1 \SI{}{\micro\meter\squared} & 30 & No & 500--630 nm & 15 nm & $\sim$1.5 \SI{}{\milli\second} & NA & Surface \\
\hline
Polychromat on CCD \cite{Wang2014broabandDiffractiveSpectrometer} & 2014 & NA & NA & No & 400--900 nm & $\sim$1 nm & NA & NA & Surface \\
\hline
PMMA etalon array on CCD \cite{Huang2017etalonArraySpectrometry} & 2017 & 5$\times$5 \SI{}{\milli\meter\squared} & 100 & No & 400--900 nm & 8 nm & NA & NA & Surface \\
\hline
Perovskite quantum dot on CCD \cite{Zhu2020perovskiteQuantumDotSpectrometer} & 2020 & 7$\times$7 \SI{}{\centi\meter\squared} & 361 & No & 250--1000 nm & $\sim$1.6 nm & NA & NA & Surface \\
\hline
Plasmonic encoder on CMOS imager \cite{Brown2021plasmonicEncoder} & 2021 & 3.6$\times$4.8 \SI{}{\milli\meter\squared} & 252 & Yes & 480--750 nm & NA & $\sim$28 \SI{}{\micro\second} & NA & Surface \\
\hline
Si nanowire detector array \cite{Meng2020structurallyColoredSiNanoWires} & 2020 & 7.7$\times$7.7 \SI{}{\milli\meter\squared} & 24 & Yes & 450--800 nm & NA & NA & NA & Surface \\
\hline
\textbf{This work: PTST enhanced Si PD array} & 2025 & 0.4 \SI{}{\milli\meter\squared} & 16 & Yes & 640--1100 nm & 8 nm & 57 \SI{}{ps} & $>$30 dB ($>$40 dB) & Surface \\
\hline
\end{tabular}
\label{tab:benchmarking_spectrometers}
\end{sidewaystable}

\subsection{Benchmarking}\label{benchmarking}

We present a benchmarking table (Table~\ref{tab:benchmarking_spectrometers}) comparing the performance of the PTST spectrometer with other computational reconstructive spectrometers. Researchers have strived to develop compact, low-cost, and high-performance spectrometers using a combination of photonic structures, nanomaterials, and quantum dots. Some works focused on achieving high spectral resolution of $\sim$1 nm \cite{Wang2014broabandDiffractiveSpectrometer,Zhu2020perovskiteQuantumDotSpectrometer}, while others focused on miniaturization \cite{Yang2019SingleNanowireSpectrometers,Cui2025tunableOptoelectronicInterface}. 
Filter-based spectrometers are simpler in design that takes advantage of the existing detector arrays like CMOS sensors and CCDs \cite{You2024CMOScompatibleFabryPerotSpectrometers,Wang2019SingleShotSpectralSensors,Huang2017etalonArraySpectrometry,Zhu2020perovskiteQuantumDotSpectrometer,Zhang2024plasmonicNanoFilterSpectrometer}. But they are limited by the poor sensitivity of the detectors in the NIR region. 
Researchers have also developed waveguide-based spectrometers in silicon platform for IR spectroscopy \cite{Li2021waveguideFilterSpectrometers,Amin2024multiModeInterferenceWaveguideSpectrometer,Cui2025tunableOptoelectronicInterface} that can achieve high spectral resolution on a relatively narrow spectral range.
Our PTST spectrometer achieves a broad spectral range of \SIrange{640}{1100}{\nano\meter} with a spectral resolution of \SI{8}{\nano\meter} using fully integrated CMOS-compatible technology. Since each detector in the spectrometer is designed to operate at high bandwidth, such as the \SI{57}{\pico\second} time response measured in our device, the overall system supports ultrahigh-speed operation. Additionally, we demonstrated a high SNR of \SI{30}{\decibel}, even in noisy conditions.

\section{Methods}\label{Methods}

The spectrometer-on-a-chip is composed of a set of unique silicon PDs with integrated PTST designed to enhance light absorption and improve spectral resolution. The PDs are fabricated on a silicon-on-insulator (SOI) substrate with an epitaxially grown silicon layer of \SI{3.2}{\micro\meter} with ~P$^+$~--~$\pi$~--~P~--~N$^+$ doping profile. The active region of the PDs is approximately \SI{1}{\micro\meter} thick, allowing ultrafast response times. The fabrication process starts with the etching of the PTST on the silicon surface, followed by the mesa etching, sidewall passivation, and metallization. The details of the fabrication process can be found in the Ref.~\cite{amita2023DesignAndFabrication} and the flow diagram is shown in the Supplementary Information Fig. \ref{supp-supfig:ptst_fabrication}. 
The PTST are designed to enhance light absorption in specific wavelengths. To accomplish this, we varied the diameter and periodicity of the PTST, in this case micro/nano holes. This allows us to achieve unique spectral responses for each PD while improving the overall responsivity of the PDs, specifically in the NIR range. Further details on the design of the PTST for hyperspectral detectors is available in the Supplementary Information Sec. \ref{supp-PTST_design}. 
The external quantum efficiency (EQE) of the PDs is measured using a calibrated light source of \SI{10}{\micro\watt} optical power. The EQE is calculated by measuring the photocurrent generated by the PDs under illumination and normalizing it with the incident optical power using equation \ref{eqn:EQE}. 
    
\begin{equation}
    EQE = \frac{I_{ph}}{P_{opt}} \cdot \frac{hc}{q\lambda}
    \label{eqn:EQE}
\end{equation}
    
where $I_{ph}$ is the photocurrent generated by the PDs, $P_{opt}$ is the incident optical power, $q$ is the charge of an electron, $h$ is Planck's constant, $c$ is the speed of light, and $\lambda$ is the wavelength of the incident light. The EQE is measured across a wavelength range of \SIrange{640}{1100}{\nano\meter}. 

The PDs are characterized using a supercontinuum laser source (NKT SuperK Extreme EXR-04) coupled with an opto-acoustic filter (NKT Select NIR). The laser source is used to illuminate the PDs through a tapered fiber tip, and the generated photocurrent is measured using Agilent 4156C parameter analyzer. The EQE of the PDs is obtained by varying the wavelength of the laser source and recording the corresponding photocurrent.
    
The spectral reconstruction process is performed using a fully connected neural network. The input to the neural network consists of the signals from the PDs, while the output layer represents the reconstructed spectral information. The network is trained using synthetic dataset of Gaussian spectra with varying peaks and spectral widths, and their corresponding PD signals calculated using the measured EQE of the PDs with the help of Eq. \ref{eqn:Iphoto}.
    
\begin{equation}
    I_{photo} = \int_{\lambda_{1}}^{\lambda_{2}} R(\lambda) \cdot S(\lambda) d\lambda
    \label{eqn:Iphoto}
\end{equation}
where $I_{photo}$ is the photocurrent generated by the PDs, $R(\lambda)$ is the external quantum efficiency of the PDs at wavelength $\lambda$, and $S(\lambda)$ is the incident light spectra as a function of wavelength $\lambda$. $\lambda_{1}$ and $\lambda_{2}$ are the lower and upper limits of the wavelength range of interest. The dataset consists of over 500,000 samples with varying spectral profiles and their corresponding photocurrents. The neural network is trained using this dataset to learn the relationship between the input signals from the PDs and the output spectral information.
    
The training process involves minimizing the loss function over several iterations of back propagation. Several loss functions have been used to achieve the best fit for sharp spectral peaks, and best fit is obtained for a combination of RMSE and the Pearson's correlation coefficient (R) between the predicted and actual spectra. The loss function is defined as follows:
\begin{equation}
    Loss = \alpha \cdot RMSE + (1 - R)
    \label{eqn:Loss}
\end{equation}
where $RMSE$ is the root mean squared error between the predicted and actual spectra, $R$ is the Pearson correlation coefficient, and $\alpha$ controls the trade-off between the two terms. For this exercise the value of $\alpha$ is considered unity. The RMSE is a good measure for the deviation from the ground truth, however, for the case of narrow width spectra the RMSE value does not provide good distinction between weak and good estimation. Contrary to that, Pearson correlation coefficient provides much higher contrast and therefore is suitable for spectra with sharp features. A comparison of the model training with different loss functions is provided in the Supplementary Information Tab. \ref{supp-tab:loss_function_comparison}.

The neural network is implemented using pytorch with the Adam optimizer used for training. The model architecture consists of 4 deep layers with ReLU activation functions, followed by a final output layer with a linear activation function. The number of neurons in each layer is adjusted based on the complexity of the spectral data. The training process is performed using a batch size of 32 and a learning rate of 0.001. The model is trained for 1000 epochs, with early stopping implemented to prevent overfitting. The learning rate was reduced to 0.0001 after 600 epochs for better convergence. The performance of the model is evaluated using a separate validation dataset, and the loss function defined in equation \ref{eqn:Loss} is used as the evaluation metric.

For the noise tolerance analysis, single peak laser spectra was used to calculate the SNR of the conventional and PTST spectrometers.
The SNR of the spectrometer is calculated using the following equation:
\begin{equation}
    SNR = 10 \cdot \log_{10} \left( \frac{P_{signal}}{P_{noise}} \right)
    \label{eqn:SNR}
\end{equation}
where $P_{signal}$ is the optical power of the signal and $P_{noise}$ is the optical power of the noise. The signal power is decoupled from the noise power using wavelet-based signal denoising technique \cite{Abramovich2006AdaptingToUnknownSparsity_MatlabWdenoise}.The noise power is then calculated by taking the standard deviation of the noise signal.
The SNR is calculated for each laser peak spectra measured in the conventional spectrometer and their respective reconstructed spectra from the PTST spectrometer. This SNR was used to evaluate the sensitivity and noise tolerance of the PTST spectrometer.

\section{Conclusion}
\label{Conclusion}

We have demonstrated a compact, silicon-based reconstructive spectrometer-on-a-chip that is fully compatible with CMOS processes, leveraging uniquely engineered silicon photodetectors (PDs) integrated with photon-trapping surface textures (PTST). The carefully designed PTST structures enhance responsivity at selective wavelengths, enabling distinctive spectral signatures across detectors. To the best of our knowledge, this represents the first monolithic silicon spectrometer of its kind to achieve CMOS integration while delivering high spectral accuracy ($>$95\%) and resolution (\SI{8}{\nano\meter}) across a broad wavelength range of \SIrange{640}{1100}{\nano\meter}.

Compared to conventional silicon spectrometers, the PTST-based design demonstrates significantly improved sensitivity and noise performance, achieving improvements in SNR exceeding \SI{30}{\decibel} at longer wavelengths. The system also maintains stable performance even under additional noise levels up to \SI{40}{\decibel}. Our results show that increasing the number of distinct detectors can further enhance reconstruction accuracy, spectral resolution, and robustness to noise. Notably, with only 16 PDs, the system achieves over 99\% peak accuracy and a correlation greater than 0.8 with the ground truth.

The AI-augmented reconstructive spectrometer also benefits from advanced PD characteristics including low breakdown voltage ($\sim$\SI{8}{\volt}), ultrafast response time (\SI{57}{\pico\second}), and high internal gain ($>$7000), making it a strong candidate for applications such as fluorescence lifetime imaging, real-time surgical guidance, and biomedical diagnostics. The integration of AI-augmentation not only improves reconstruction accuracy but also contributes to robust noise handling, enabling performance beyond the physical limits of conventional silicon detectors.

Future work will explore the operation of PTST-PDs in the avalanche breakdown regime to further boost responsivity and sensitivity. Overall, this spectrometer-on-a-chip presents a compelling solution for compact, low-cost, and high-performance hyperspectral imaging systems, with strong potential for integration into portable and field-deployable platforms for environmental monitoring, remote sensing, precision agriculture, and consumer electronics.

% \disclosures 
\section*{Disclosures}
The authors declare that there are no financial interests, commercial affiliations, or other potential conflicts of interest that could have influenced the objectivity of this research or the writing of this paper.

\section* {Code, Data, and Materials Availability} 
The data used in this manuscript is available upon request.

\section* {Acknowledgments}
This work was supported in part by the National Institute of Biomedical Imaging and Bioengineering (NIN-NIBIB) grant \#1-P41-EB032840-01, UC Davis and LBNL Partnership Faculty Fellows Award, and by the National Science Foundation PFI-TT Award \#2329884. The authors acknowledge the Center for Nano and Micro Manufacturing (CNM2) for their support in the device fabrication process.

%%%%% References %%%%%

\bibliography{bibliography}   % bibliography data in report.bib
\bibliographystyle{spiejour}   % makes bibtex use spiejour.bst

%%%%% Biographies of authors %%%%%

\vspace{2ex}\noindent\textbf{Ahasan Ahamed, Ph.D.,} is a postdoctoral scholar at University of California, Davis. He completed his PhD degree in electrical and electronics engineering from the UC Davis focusing on on-chip spectroscopy using photon-trapping silicon photodiodes. He completed his M.Sc. degree from UC Davis in Electrical and Computer Engineering and his B.Sc. degree from Bangladesh University of Engineering and Technology (BUET) in Electrical and Electronics Engineering. 

His current research interests include optoelectronic sensors, silicon photonics, hyperspectral imaging, hardware security and photonic true random number generators. He received prestigious Smita Bakshi teaching award and Graduate Student Leadership award for his contribution to the academic and social community. He is actively participating outreach activities to promote STEM education through CITRIS-INSPIRE and GREAT program. He is a current member of SPIE and co-chaired a session in low-dimensional materials and devices in SPIE Optics and Photonics conference. He has authored and co-authored over 20 research papers in peer-reviewed journals and conferences.

\vspace{2ex}\noindent\textbf{Htet Myat} is a research intern at the University of California, Davis. He received a Bachelor of Science in Computer Engineering from the University of California, Davis, in 2023. Their research interests include the integration of machine learning and embedded system engineering.

\vspace{2ex}\noindent\textbf{Amita Rawat, Ph.D.,} did her undergrad studies at the Department of Electrical Engineering, IIT Patna, India, in 2013. After graduation, she worked for two years on SRAM cell designing. She started her Ph.D. in 2015 in the Department of Electrical Engineering at IIT Bombay, India. During her Ph.D., she acquired expertise in semiconductor devices. She worked on process-induced variability modeling for advanced logic devices such as FinFETs, Nanowire FETs, and Nanosheet FETs. She also filed a patent securing the method to develop a low-cost SOI wafer. She graduated with an Excellence in Research award in 2021. 

After her Ph.D., she worked as a PDK development engineer at IMEC, Leuven, Belgium. In collaboration with Huawei, she patented a novel Nanosheet device architecture essential for restoring the lost strain during the source/drain regrowth process. After a year of experience, she moved to the United States and assumed a post-doctoral fellowship position at the University of California, Davis. As a postdoc, her research area focuses on designing and fabricating high-speed photodetectors for biomedical imaging, LiDAR, and data-communication applications. At UC Davis, she has volunteered to teach as a guest lecturer, served as a reviewer for a postdoctoral research symposium, and played a key role in organizing semiconductor device fabrication training workshops for high school and undergraduate students. Insofar, she filed two patents and published 21 research papers.

\vspace{2ex}\noindent\textbf{Lisa N. McPhillips} is a PhD student in electrical and computer engineering at the University of California, Davis. She completed her undergraduate degree in bioengineering at the University of California, Santa Cruz in 2020. 

\vspace{2ex}\noindent\textbf{M. Saif Islam, Ph.D.,} is a professor and chair of Electrical and Computer Engineering at the University of California, Davis. He earned his B.Sc. in Physics from Middle East Technical University, M.Sc. in Physics from Bilkent University, and both M.S. and Ph.D. degrees in Electrical Engineering from UCLA. He served as the director of the Center for Information Technology Research in the Interest of Society (CITRIS) and the Banatao Institute. 

Dr. Islam's research focuses on nanotechnology, particularly the integration of low-dimensional and nanostructured materials with conventional semiconductor systems. His work spans applications in nanoelectronics, ultrafast optoelectronics, quantum sensing, energy harvesting, and computational imaging. He has authored over 250 scientific publications, holds 43 patents, and has co-founded two startups based on his inventions.

A fellow of IEEE, AAAS, OSA, SPIE, and the National Academy of Inventors, Dr. Islam is also a recipient of the NSF CAREER Award and UC Davis's Academic Senate Distinguished Teaching Award—the university's highest teaching honor.

% \vspace{1ex}
% \noindent Biographies and photographs of the other authors are not available.

\listoffigures
\listoftables

\clearpage
\section*{Supplementary Information}
% \documentclass[12pt]{spieman}  % 12pt font required by SPIE;
% \documentclass[a4paper,12pt]{spieman}  % use this instead for A4 paper
% \usepackage{amsmath,amsfonts,amssymb}
% \usepackage{graphicx}
% \usepackage{setspace}
% \usepackage{tocloft}
% \usepackage{lineno}
% \usepackage{siunitx}
% \usepackage{rotating}
% % \usepackage[labelfont=bf]{caption} % Makes all figure labels bold

% \linenumbers
\title{
    AI-Augmented Photon-Trapping Spectrometer-on-a-Chip on Silicon Platform with Extended Near-Infrared Sensitivity
    % Photon-Trapping Silicon Photodiodes Enable Noise-Tolerant Spectrometer-on-a-Chip for Hyperspectral Imaging with Extended Sensitivity up to 1100 nm
}

\author[a]{Ahasan Ahamed}
\author[a]{Amita Rawat}
\author[a]{Htet Myat}
\author[a]{Lisa N McPhillips}
\author[a*]{M Saif Islam}
\affil[]{University of California Davis, Electrical and Computer Engineering Department, One Shields Avenue, Davis, CA 95616, USA}

\renewcommand{\cftdotsep}{\cftnodots}
\renewcommand{\thefigure}{S\arabic{figure}}
\cftpagenumbersoff{figure}
\cftpagenumbersoff{table} 
% \begin{document} 
\maketitle

% Include email contact information for corresponding author
{\noindent \footnotesize\textbf{*}M Saif Islam,  \linkable{sislam@ucdavis.edu} }

\begin{spacing}{2}   % use double spacing for rest of manuscript
%% \section{PTST Design}
%% \section{Fabrication}
%% \section{EQE}
%% \section{Reconstruction - IV} Why less sensitivity still works
%% \section{Loss Function Comparison}
%% \section{Spectrometer Performance}
%% \section{Simulated EQE for HSIC}

\section{PTST Design}
\label{PTST_design}    % \label{} allows reference to this section
\begin{figure}[htbp]
\centering
\includegraphics[width=0.6\textwidth]{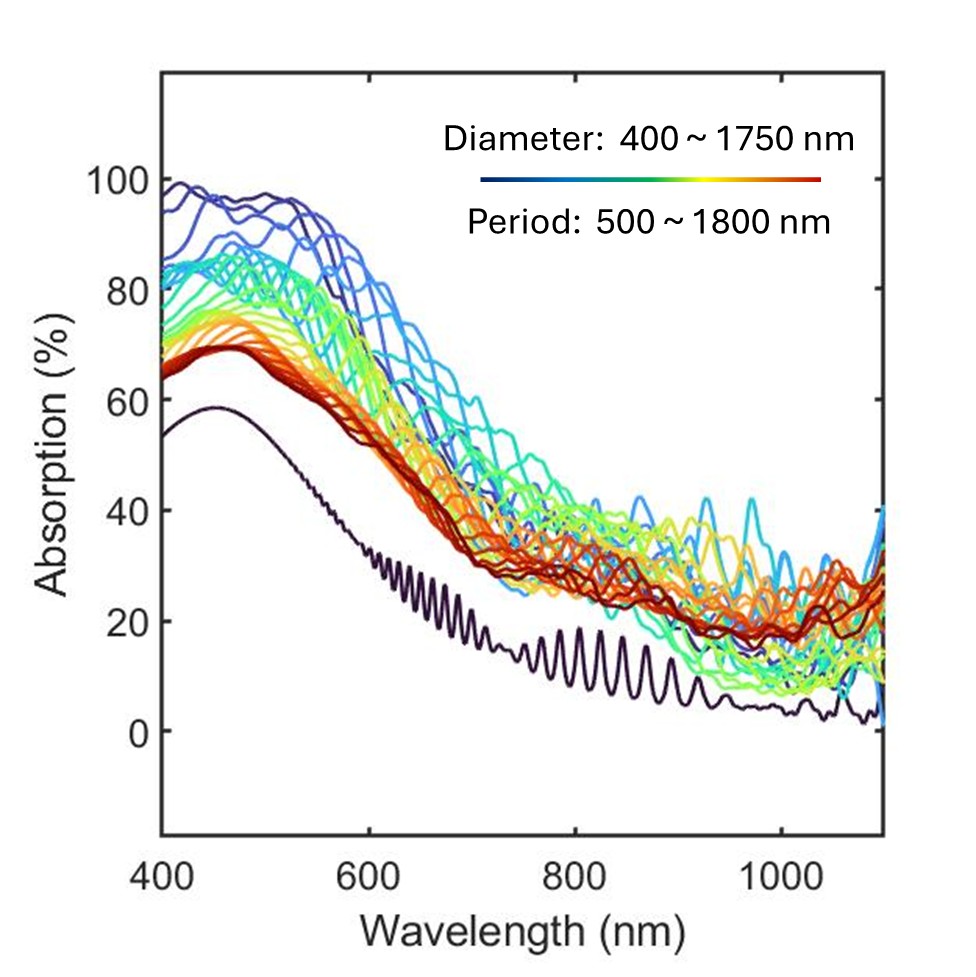}
\caption[Simulated absorption spectra of different PTST]{
Simulated absorption spectra of different PTST: The simulated absorption profile of the silicon PDs with varying periodicities of PTST ranging from \SIrange{400}{1800}{\nano\meter}. The peak absorption wavelength shifts to longer wavelengths with increasing periodicity of the PTST, allowing us to achieve unique spectral response across the broad spectral range of \SIrange{400}{1100}{\nano\meter}. 
}
\label{supfig:simulated_eqe}
\end{figure}

Recent works on photon-trapping structures (PTST) demonstrated that the peak absorption wavelength is directly proportional to the periodicity of the nanostructures \cite{ahasan2023UniqueHyperspectralResponse}. Therefore, we designed a set of PTST that varies in periodicity to achieve unique spectral response across the broad spectral range of \SIrange{400}{1100}{\nano\meter}. To accomplish that, we varied the periodicity of the PTST from \SIrange{500}{1800}{\nano\meter} in steps of \SI{50}{\nano\meter} with a minimum hole-to-hole spacing of \SI{50}{\nano\meter}. The diameter on the PTSTs are varied from \SIrange{400}{1750}{\nano\meter} These PTST structures are then simulated using finite distance time domain (FDTD) method to calculate the absorption profile of the silicon PDs. The simulated absorption profile is shown in Fig.~\ref{supfig:simulated_eqe} for different periodicities of PTST. The absorption profile shows that the peak absorption wavelength shifts to longer wavelengths with increasing periodicity of the PTST. This allows us to achieve unique spectral response across the broad spectral range of \SIrange{400}{1100}{\nano\meter}. The simulated absorption profile is then used to design the PTST structures on top of the silicon PDs.
However, due to practical limitations, only a subset of the designed PTST were fabricated. 

\section{Device Fabrication}
\label{Device_Fabrication}

\begin{figure}[htbp]
\centering
\includegraphics[width=0.9\textwidth]{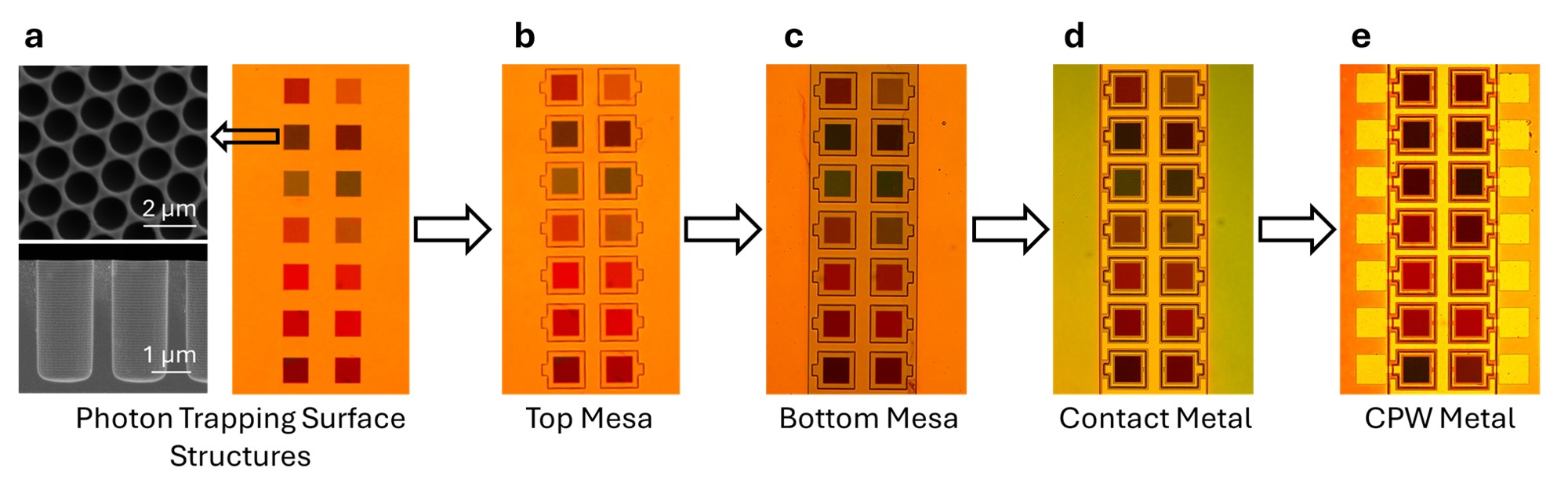}
\caption[Device fabrication]{Device fabrication: (a) Microscope image of the PTST patterns transferred onto the silicon layer using DRIE. Each PTST pattern is unique and optimized for specific wavelengths. This can readily observed from the structurally colored properties of the PTST patterns. On the left side, the SEM images show (top) the top view and (bottom) the cross sectional view of the PTST patterns. (b) Microscope image of the top mesa etched using DRIE.
The top mesa is etched to a depth of $\sim$\SI{2.3}{\micro\meter} on the P-on-N doping profile. The etch depth is critical for proper contact to the highly doped region at the bottom of the photodiode. (c) Microscope image of the bottom mesa etched using DRIE. The bottom mesa is etched to the buried oxide layer to completely isolate the photodiode from the substrate. HF passivation is done to remove the dangling bonds from the PTST and sidewall which drastically reduces the dark current. (d) Microscope image of the aluminum contact metal pads deposited using electron beam evaporation and patterned using photolithography and lift-off process. The contact metal pads are used to make electrical connections to the photodiode. (e) Microscope image of the completed device with coplanar waveguide (CPW) contact metal deposited using electron beam evaporation and patterned using photolithography and lift-off process. This step was done after the deposition and etch of sidewall oxide.}
\label{supfig:ptst_fabrication}
\end{figure}

The fabrication of the PTST spectrometer-on-a-chip is performed using a CMOS compatible process. The fabrication process starts with the epitaxial growth of silicon on a silicon-on-insulator (SOI) wafer. The doping profile consist of a sequential ~P$^+$~--~$\pi$~--~P~--~N$^+$ layers to achieve low breakdown voltage and high gain \cite{amita2023DesignAndFabrication}. The fine PTST designs are patterned using a deep ultraviolet (DUV) ASML lithography tool with a 4$\times$ magnification. This allowed us to achieve a minimum feature size of \SI{300}{\nano\meter}. Due to this practical limitation the PTST design in the fabricated devices have periodicities from \SIrange{800}{1600}{\nano\meter}, therefore they are optimized for near infrared wavelengths. The PTST patterns are designed to be unique which can be readily observed from the structurally colored properties of the PTST patterns under the microscope (Fig.~\ref{supfig:ptst_fabrication}(a)). 
The PTST patterns are then transferred onto the silicon layer using deep reactive ion etching (DRIE) process. The PTST patterns are designed to be unique and optimized for specific wavelengths, which can be readily observed from the structurally colored properties of the PTST patterns. The top mesa is etched to a depth of $\sim$\SI{2.3}{\micro\meter} to reach to the bottom $N+$ contact of the photodiode. The bottom mesa is etched up to the buried oxide layer to completely isolate the photodiode from the substrate. Next, a diluted hydrofluoric acid (HF) was used to passivate the dangling bonds from the etched PTST and sidewall, otherwise this can lead to drastical increase in the dark current \cite{Gao2017photonTrappingMicrostructures}. Finally, aluminum contact metal pads are deposited using electron beam evaporation and patterned using photolithography and lift-off process. The completed device with the gold dposited coplanar waveguide (CPW). The process flow is illustrated in Fig.~\ref{supfig:ptst_fabrication}.

\section{EQE of PTST Photodiodes}
\label{EQE_PTST}
\begin{figure}
\centering
\includegraphics[width=1\textwidth]{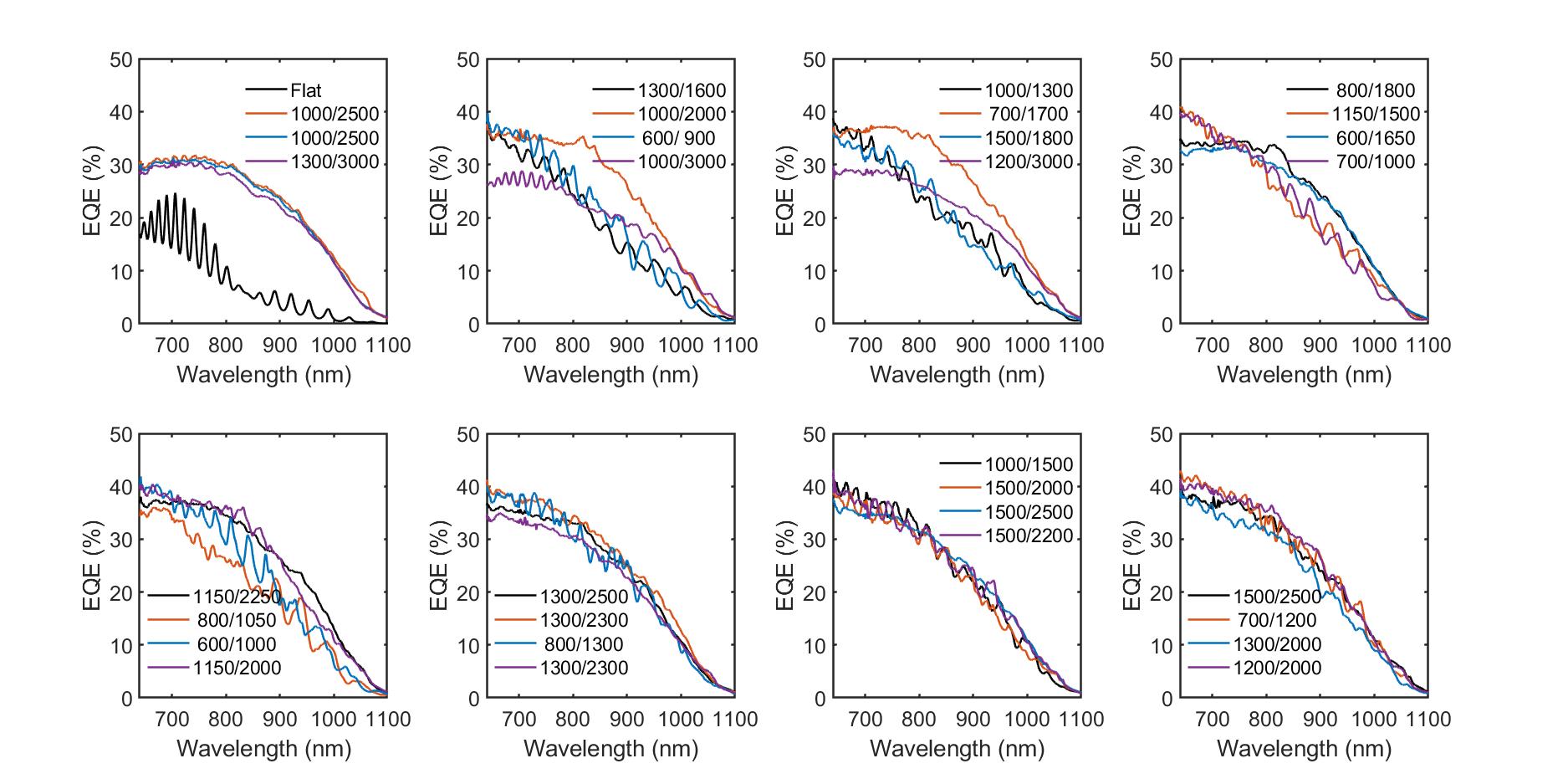}
\caption[EQE spectra of PTST photodiodes]{EQE spectra of PTST photodiodes: The EQE spectra corresponding to individual PTST structures. The diameter and periodicity of the PTST structures are mentioned in the legend in nanometers.}
\label{supfig:eqe_spectra}
\end{figure}

The external quantum efficiency (EQE) of the PTST photodiodes is calculated from the measured photocurrent and the incident optical power. The photocurrent is measured using Agilent 4156C precision semiconductor parameter analyzer. The incident optical power is measured using a Thorlabs S130VC power meter. NKT SuperK Extreme EXR-4 supercontinuum laser is used as the primary light source which was tuned to single wavelengths in the range from \SIrange{640}{1100}{\nano\meter} using NKT Select NIR module. The dark current of the PDs are subtracted from the measured photocurrent to obtain the net photocurrent. The EQE spectra corresponding to individual PTST structures is shown in Fig.~\ref{supfig:eqe_spectra}. 

\section{Neural Network Model}
\label{Neural_Network_Model}
The neural network model is trained on a synthetic dataset generated using the measured EQE spectra of the PTST enabled PDs. PDs that can reproduce the same spectral response and generate the same photocurrent for the same laser illumination are considered for the training, therefore any PDs with defects are excluded. The photocurrent from the measurement and calculated EQE spectra are compared to make sure that the synthetic dataset is representative of the actual spectral response of the PTST enabled PDs. 

The neural network model contains one input layer with input number of PDs, 4 hidden layers with 1024 neurons each, and one output layer of 461 neurons corresponding to the number of wavelengths. We used several loss functions to train the model, and the best performance was achieved using a combination of root mean square error (RMSE) and Pearson correlation coefficient (r). The loss function is defined as:
\begin{equation}
    \text{Loss} = \text{RMSE} + (1 - r)
\end{equation}
where RMSE is the root mean square error between the predicted and actual spectral profile, and r is the Pearson correlation coefficient between the predicted and actual spectral profile.

The other loss functions used for training the model are RMSE, Pearson correlation coefficient (r), coefficient of determination ($R^2$), and dot product similarity. The table \ref{tab:loss_function_comparison} shows the performance of the neural network model for different loss functions. 

\begin{table}
\centering
\caption[Loss function comparison]{Loss function comparison: The performance of the neural network model for different loss functions.}
\begin{tabular}{|m{3.8cm}|>{\centering\arraybackslash}m{1.8cm}|>{\centering\arraybackslash}m{2.1cm}|>{\centering\arraybackslash}m{1.6cm}|m{4.6cm}|}
\hline
Loss Function & RMSE & RMSE & Pearson r & Comment \\
& (synthetic) &(experiment) & (average) & \\
\hline
RMSE & 0.03019 & 0.05188 & 0.8574 & Poor at resolving sharp peak shapes \\
Pearson Correlation Coefficient (r) & 0.03857 & 0.05827 & 0.84066 & Cannot predict amplitude properly \\
Dot Product Similarity & 0.04215 & 0.06426 & 0.8194 & Cannot predict amplitude properly \\
\textbf{RMSE + (1 - r)} & \textbf{0.02868} & \textbf{0.04864} & \textbf{0.86545} & Can predict amplitude and shape properly \\
\hline
\end{tabular}
\label{tab:loss_function_comparison}
\end{table}

Although RMSE is best suited for most regression problems \cite{ahasan2024OnChipHyperspectralDetectors}, in this case, it does not work well since the laser spectra are narrow peaks in a wide spectral range. This means that the RMSE is dominated by the large number of zero values in the spectral profile, leading to a poor performance. While the Pearson correlation coefficient (r) can capture the shape of the spectral profile, it does not take into account the magnitude of the spectral profile. Similarly dot product similarity, and coefficient of determination ($R^2$) are also not suitable for this problem. Therefore, we used a combination of RMSE and Pearson correlation coefficient (r) as the loss function to train the neural network model. This allows us to achieve a good balance between the magnitude and shape of the spectral profile that works well for narrow peaks as well as broad spectral profiles.

\section{Unique Response \& Sensitivity}
\label{Unique_Response}
\begin{figure}
\centering
\includegraphics[width=1\textwidth]{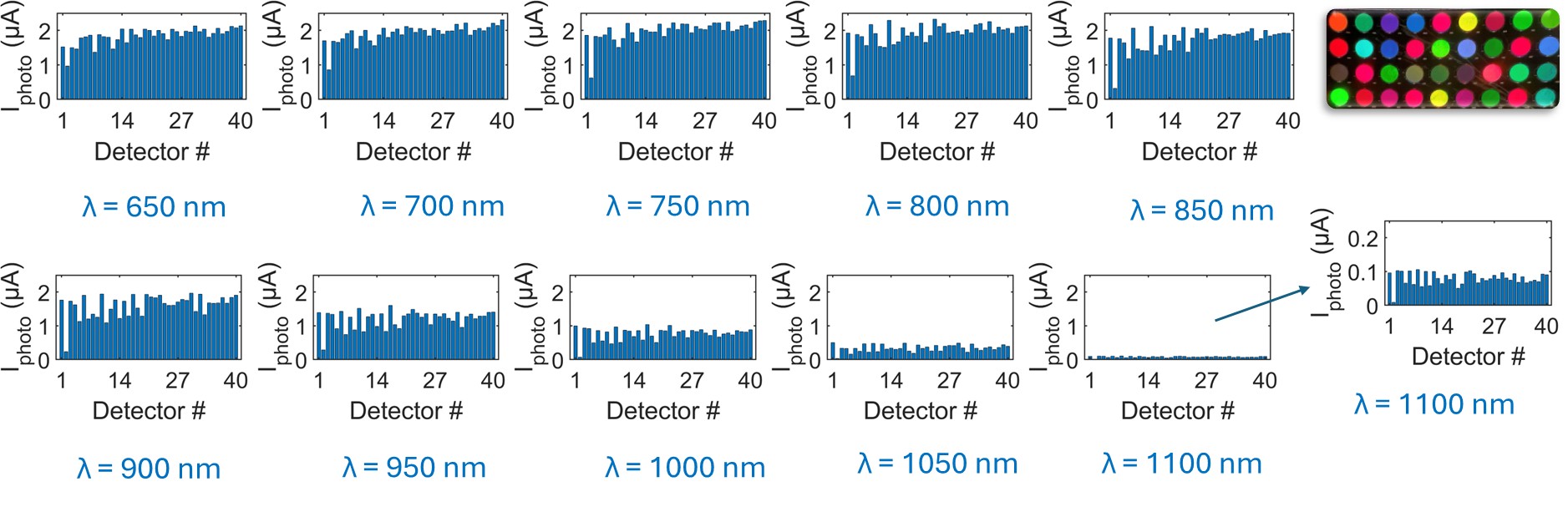}
\caption[Measured photocurrent of PTST photodiodes]{Measured photocurrent of PTST photodiodes: The measured photocurrent of all fabricated PDs for laser illumination ranging from \SIrange{650}{1100}{\nano\meter} with laser power of \SI{10}{\micro\watt}. The photocurrent across various PDs vary significantly due to the unique spectral response of the PTST. This allows us to reconstruct spectral profile at longer wavelengths despite the reduced EQE in silicon PDs. The optical image of the PDs is shown in the top right corner.}
\label{supfig:unique_response}
\end{figure}

Although the silicon PDs have reduced EQE at longer wavelengths, the PTST spectrometer is capable of detecting the laser peaks even at \SI{1100}{\nano\meter} due to the unique spectral response of the PTST structures. To better explain this we plotted the measured photocurrents of all fabricated PDs in Fig.~\ref{supfig:unique_response} for laser illumination ranging from \SIrange{650}{1100}{\nano\meter}. The laser power is calibrated to remain at \SI{10}{\micro\watt}. The photocurrent across different PDs vary from each other for different wavelengths due to the presence of unique PTST in the fabricated PDs. This variation in the photocurrent allows the neural network to differentiate the spectral profile of the laser peaks for different wavelengths. Also, since the IV profile is unique for each wavelength, it is possible to identify the laser peaks even at longer wavelengths without the need for high EQE. 

\section{Spectral resolution}
\label{Spectral_Resolution}
\begin{figure}
\centering
\includegraphics[width=1\textwidth]{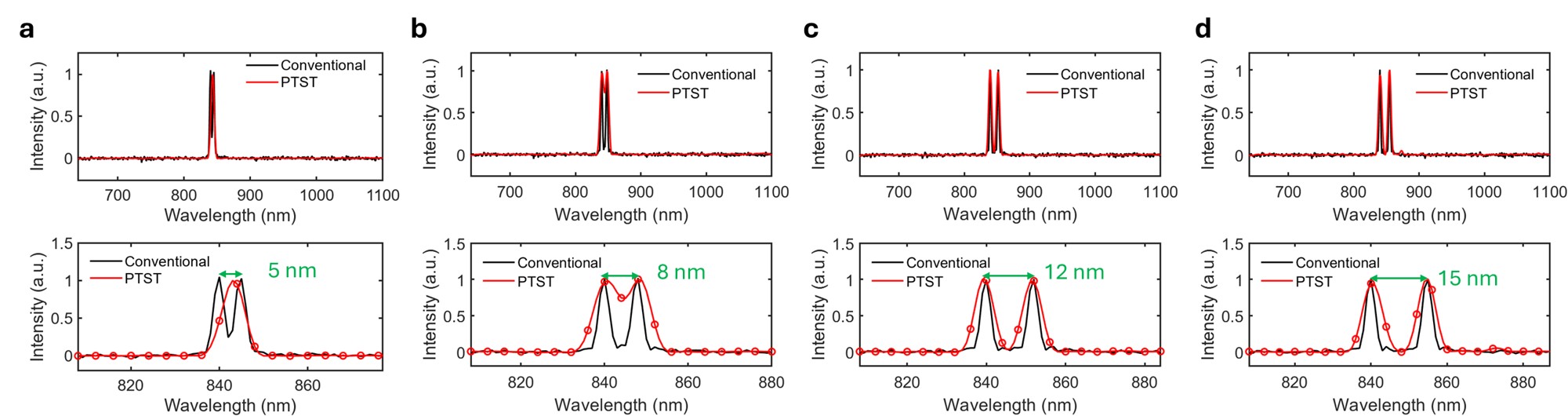}
\caption[Spectral resolution of the PTST spectrometer]{Spectral resolution of the PTST spectrometer: The spectral reconstruction of the laser spectra separated by (a) \SI{5}{\nano\meter}, (b) \SI{8}{\nano\meter}, (c) \SI{12}{\nano\meter}, and (d) \SI{15}{\nano\meter} is performed using the neural network model with 32 photodiodes. The reconstructed spectra can distinguish closely spaced laser spectra up to 8 nm, demonstrating high spectral resolution of the spectrometer-on-a-chip. For \SI{5}{\nano\meter} separation in (a) the reconstructed spectra is merged and the spectrometer cannot
 differentiate the individual peaks.}
\label{supfig:spectral_resolution}
\end{figure}

The spectral resolution defined as the minimum separation between two peaks that can be resolved by the spectrometer. To determine the spectral resolution of the PTST spectrometer, we illuminated the device with two laser peaks at varying wavelength separations and measured the photocurrent from 32 PDs, then the photocurrent was used to reconstruct the spectral profile using the trained neural network. In Fig.~\ref{supfig:spectral_resolution} we show the spectral reconstruction of the laser spectra separated by \SI{5}{\nano\meter}, \SI{8}{\nano\meter}, \SI{12}{\nano\meter}, and \SI{15}{\nano\meter}. The PTST spectrometer is capable of resolving laser spectra with a separation of \SI{8}{\nano\meter} or more. For \SI{5}{\nano\meter} separation, the reconstructed spectra is merged and therefore the spectra cannot be resolved. To improve the spectral resolution of the PTST spectrometer, one can increase the number of unique PDs used for spectral reconstruction or improve the uniqueness in the spectral response of the PTST enhanced PDs.

\section{Dynamic Range}
\label{Dynamic_Range}
\begin{figure}
\centering
\includegraphics[width=1\textwidth]{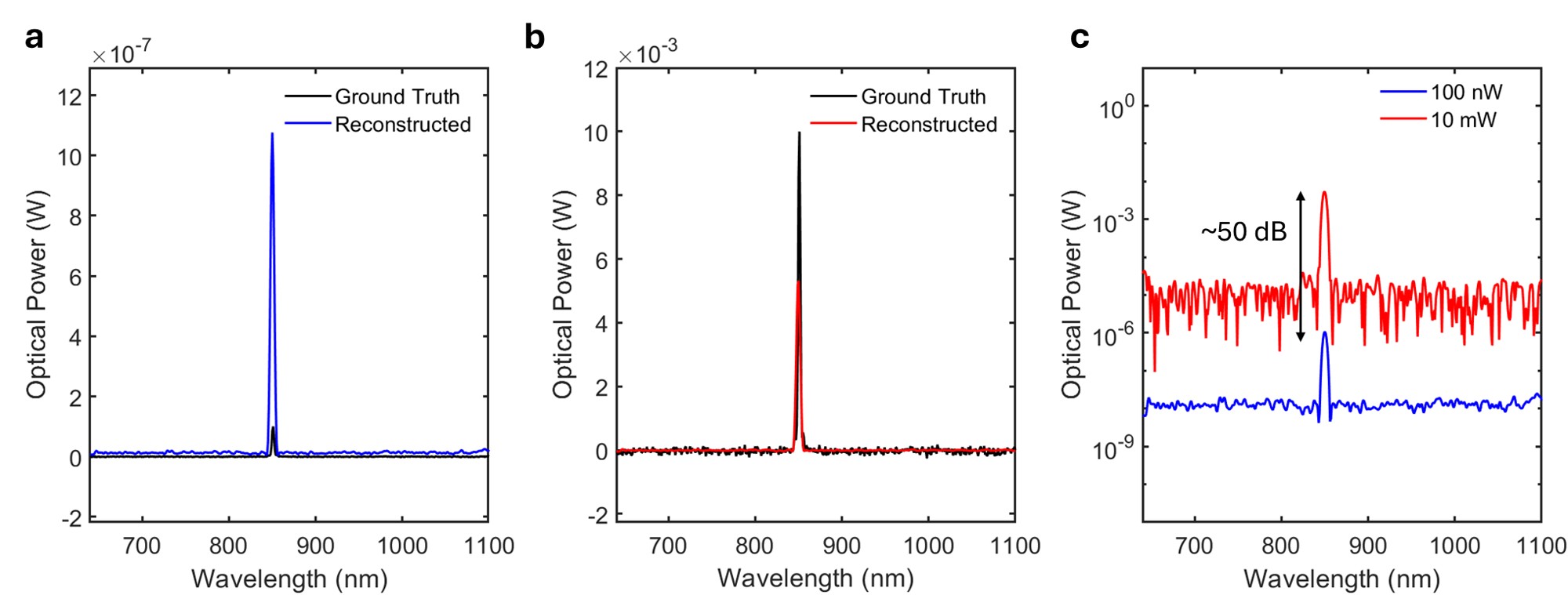}
\caption[Dynamic range of the PTST spectrometer]{Dynamic range of the PTST spectrometer: The reconstructed spectral profile for \SI{850}{\nano\meter} laser illumination at power (a) \SI{100}{\nano\watt} and (b) \SI{10}{\milli\watt} respectively. The spectrometer can detect the signal from \SI{100}{\nano\watt} to \SI{10}{\milli\watt} with a signal-to-noise ratio (SNR) of $\geq$\SI{20}{\decibel}. (c) The dynamic range of the PTST spectrometer is observed to be $\geq$\SI{50}{\decibel}.}
\label{supfig:dynamic_range}
\end{figure}

The dynamic range of the PTST spectrometer is defined as the ratio of the maximum detectable signal to the minimum detectable signal. For this measurement, we linearly increased the photocurrent from the PDs by varying the laser power from \SI{100}{\nano\watt} to \SI{10}{\milli\watt} at a fixed wavelength of \SI{850}{\nano\meter}. The reconstructed spectral profile is shown in Fig.~\ref{supfig:dynamic_range} for different laser powers. The spectrometer can detect the signal down to \SI{100}{\nano\watt} with a signal-to-noise ratio (SNR) of $\sim$\SI{20}{\decibel}. The maximum detectable signal is limited by laser power. The dynamic range of the PTST spectrometer is calculated to be $\sim$\SI{50}{\decibel}. Note that the reconstructed spectral profile at higher laser power is not linear, as the neural network is trained on normalized spectra with approximately \SI{10}{\micro\watt} power level. 

% \disclosures 
% \section*{Disclosures}
% The authors declare that there are no financial interests, commercial affiliations, or other potential conflicts of interest that could have influenced the objectivity of this research or the writing of this paper.

% \section* {Code, Data, and Materials Availability} 
% The data used in this manuscript is available upon request.

% \section* {Acknowledgments}
% The authors would like to acknowledge the support from the National Science Foundation (NSF) and National Institute of Health (NIH). The authors acknowledge the Center for nano and micro manufacturing (CNM2) for their support in device fabrication process. The authors also thank the University of California Davis for providing the necessary facilities and resources for this research.

%%%%% References %%%%%

% \bibliography{bibliography}   % bibliography data in report.bib
% \bibliographystyle{spiejour}   % makes bibtex use spiejour.bst

%%%%% Biographies of authors %%%%%

% \vspace{2ex}\noindent\textbf{Ahasan Ahamed} is a postdoctoral scholar at University of California Davis. He completed his PhD degree in electrical and electronics engineering  from the University of California Davis focusing on on-chip spectroscopy using photon-trapping silicon photodiodes. His current research interests include optoelectronic sensors, silicon photonics, hyperspectral imaging, and photonic chips. He is an active member of SPIE.

% \vspace{1ex}
% \noindent Biographies and photographs of the other authors are not available.

% \listoffigures
% \listoftables

\end{spacing}
% \end{document}

%% \section{PTST Design}
%% \section{Fabrication}
%% \section{EQE}
%% \section{Reconstruction - IV} Why less sensitivity still works
%% \section{Loss Function Comparison}
%% \section{Spectrometer Performance}
%% \section{Simulated EQE for HSIC}

\end{spacing}

\end{document}